\documentclass[american,reprint]{revtex4-2}
\usepackage[T1]{fontenc}
\usepackage[utf8]{inputenc}
\usepackage{babel}
\usepackage{mathrsfs}
\usepackage{bm}
\usepackage{amsmath}
\usepackage{amssymb}
\usepackage{graphicx}
\usepackage{esint}
\usepackage[pdfusetitle,
 bookmarks=true,bookmarksnumbered=false,bookmarksopen=false,
 breaklinks=false,pdfborder={0 0 1},backref=false,colorlinks=false]
 {hyperref}

\makeatletter

\newcommand{\lyxmathsym}[1]{\ifmmode\begingroup\def\b@ld{bold}
  \text{\ifx\math@version\b@ld\bfseries\fi#1}\endgroup\else#1\fi}

\providecommand{\tabularnewline}{\\}

\usepackage{mhchem}
\usepackage{bm}

\makeatother

\begin{document}
\title{No source-free exchange-correlation magnetic fields in non-collinear
spin-density functional theory}
\author{Ester Livshits}
\affiliation{Fritz Haber Center for Molecular Dynamics, Institute of Chemistry,
The Hebrew University of Jerusalem, Jerusalem 9190401, Israel}
\author{Roi Baer}
\email{corresponding author: roi.baer@gmail.com}

\affiliation{Fritz Haber Center for Molecular Dynamics, Institute of Chemistry,
The Hebrew University of Jerusalem, Jerusalem 9190401, Israel}
\begin{abstract}
The source-free condition, $\nabla\cdot\mathbf{B}_{\mathrm{xc}}=0$,
where $\mathbf{B}_{\mathrm{xc}}=\delta E_{\mathrm{xc}}/\delta\mathbf{m}$
is the exchange-correlation (xc) magnetic field of non-collinear spin-density
functional theory (SDFT) and $\mathbf{m}$ the magnetization, is widely
believed to be exact. Recent studies report that projecting the field
of a parent functional onto its source-free part improves the predicted
magnetic moments. We show that the condition violates the global spin-rotation
symmetry of $E_{\mathrm{xc}}$ and therefore cannot be exact. A source-free
field is exact only in a transverse theory whose xc functional differs
from $E_{\mathrm{xc}}$ and is itself not invariant under spin rotations.
We examine a source-free (SoF) construction, an exchange-correlation
energy given by a parent SDFT functional evaluated at the transverse
magnetization $\mathbf{m}_{\perp}$. Its field is divergence-free,
remains a functional derivative and exerts local torques, but still,
it breaks global spin-rotation symmetry. With the LSDA as parent we
examine the \ce{Mn2} molecule. The SoF and the locally collinear
(LoC) LSDA both give too short a bond and too large a bond energy,
but SoF finds a broken-symmetry antiferromagnetic solution, consistent
with the experimentally observed $^{1}\Sigma^{+}_{g}$ ground state,
with a coupling of the right sign and roughly the right size, while
LoC finds a high-spin $^{11}\Pi_{u}$ ferromagnet and the wrong sign.
The broken symmetry of SoF, however, produces severe errors in directional
properties: a rigid spin rotation of the magnetization changes the
energy by about $1\,\mathrm{eV}$, in a weak uniform field the magnetization
points perpendicular to the field instead of antiparallel to it, and
the molecule develops a spurious magnetic anisotropy.
\end{abstract}
\maketitle

\section{Introduction}

\label{sec:intro}

Molecules and solids can have local magnetic moments that point in
different directions: canted or frustrated magnets, spin spirals,
an antiferromagnet in an applied field. The magnetization is then
a vector field, and the electron spin has to be treated in full. We
do that with a many-electron Hamiltonian in which an external field
couples to the spin of each electron \footnote{For simplicity we omit the orbital coupling $\mathbf{p}_{i}\to\mathbf{p}_{i}+\mathbf{A}(\mathbf{r}_{i})/c$.
That drops diamagnetic and orbital-paramagnetic response, which we
do not need here. Those terms are spin scalars: they commute with
$\hat{\mathbf{S}}$, drop out of every derivative with respect to
a global spin rotation, and leave the Hamiltonian covariant under
that rotation.} 
\begin{equation}
\hat{H}=\sum_{i}\Bigl[-\tfrac{1}{2}\nabla^{2}_{i}+v_{\mathrm{ext}}(\mathbf{r}_{i})+\bm{\sigma}_{i}\cdot\mathbf{B}_{\mathrm{ext}}(\mathbf{r}_{i})\Bigr]+\frac{1}{2}\sum_{i\neq j}\frac{1}{|\mathbf{r}_{i}-\mathbf{r}_{j}|},\label{eq:Many-body-hamiltonian}
\end{equation}
where $\bm{\sigma}_{i}=(\sigma^{x}_{i},\sigma^{y}_{i},\sigma^{z}_{i})$
is the vector of Pauli matrices acting on the spin of electron $i$,
and $v_{\mathrm{ext}}$ and $\mathbf{B}_{\mathrm{ext}}$ are the external
scalar potential and magnetic field \footnote{We use atomic units, $\hbar=m_{e}=e=1$, in which the Bohr magneton
is $\mu_{B}=e\hbar/2m_{e}=1/2$. The Zeeman energy is usually written
as $-\int\mathbf{m}_{\mathrm{phys}}\cdot\mathbf{B}_{\mathrm{phys}}\,d^{3}r$,
where $\mathbf{m}_{\mathrm{phys}}=-g\mu_{B}\mathbf{s}$ is the magnetization
density, $g$ the Landé factor and the minus sign reflects the negative
electron charge. $\mathbf{s}(\mathbf{r})=\langle\sum_{i}\mathbf{S}_{i}\delta(\mathbf{r}-\mathbf{r}_{i})\rangle$
is the spin density with $\mathbf{S}_{i}=\bm{\sigma}_{i}/2$. We follow
the tradition of SDFT literature which customarily defines the magnetization
as $\mathbf{m}=-\mathbf{m}_{\mathrm{phys}}/\mu_{B}$, the magnetic
field as $\mathbf{B}_{\mathrm{ext}}\equiv\mu_{B}\mathbf{B}_{\mathrm{phys}}$
(the field in $E_{h}/\mu_{B}$) and approximates $g=2$, so that $\mathbf{m}=2\mathbf{s}=\langle\sum_{i}\bm{\sigma}_{i}\delta(\mathbf{r}-\mathbf{r}_{i})\rangle$
as in Eq.~(\ref{eq:The-Densities}). The Zeeman energy then takes
the form $+\int\mathbf{m}\cdot\mathbf{B}_{\mathrm{ext}}\,d^{3}r$
used in Eq.~(\ref{eq:Many-body-hamiltonian}). Note that minimizing
the energy aligns $\mathbf{m}_{\mathrm{phys}}$ with $\mathbf{B}_{\mathrm{phys}}$
and hence antialigns $\mathbf{m}$ with $\mathbf{B}_{\mathrm{ext}}$.}. We write $\hat{H}=\hat{T}+\hat{W}+\hat{V}$, with $\hat{T}$ the
kinetic energy, $\hat{W}$ the electron--electron interaction, and
$\hat{V}$ the coupling to the external potentials.

As usual, finding the many-body ground state wave function and energy
of Eq.~(\ref{eq:Many-body-hamiltonian}) directly is feasible only
for the smallest systems and a useful procedure is to opt for non-collinear
spin density functional theory (SDFT) \citep{vonBarthHedin1972,Kubler1988,Sandratskii1998},
which replaces the many-electron wave function with the particle density
$n$ and magnetization density $\mathbf{m}$ as the basic variables.
These are the ground-state expectation values
\begin{equation}
n(\mathbf{r})=\Bigl\langle\sum_{i}\delta(\mathbf{r}-\mathbf{r}_{i})\Bigr\rangle,\qquad\mathbf{m}(\mathbf{r})=\Bigl\langle\sum_{i}\bm{\sigma}_{i}\,\delta(\mathbf{r}-\mathbf{r}_{i})\Bigr\rangle.\label{eq:The-Densities}
\end{equation}

The total energy becomes a functional of this pair: 
\begin{equation}
E[n,\mathbf{m}]=F[n,\mathbf{m}]+\int\bigl[v_{\mathrm{ext}}(\mathbf{r})n(\mathbf{r})+\mathbf{B}_{\mathrm{ext}}(\mathbf{r})\cdot\mathbf{m}(\mathbf{r})\bigr]\,d^{3}r,\label{eq:total-energy-functional}
\end{equation}
where the universal functional 
\begin{equation}
F[n,\mathbf{m}]=\min_{\Gamma\to(n,\mathbf{m})}\mathrm{Tr}\bigl[(\hat{T}+\hat{W})\Gamma\bigr]\label{eq:constrained-search}
\end{equation}
is defined by constrained search over $N$-electron density matrices
$\Gamma$~\citep{Levy1979,Valone1980,Lieb1983}. We must use this
and not the Hohenberg--Kohn (HK) map because in SDFT distinct pairs
$(v_{\mathrm{ext}},\mathbf{B}_{\mathrm{ext}})$ can generate the same
ground-state densities~\citep{CapelleVignale2001,EschrigPickett2001},
so that the HK map is not invertible. The domain of $F$ is the set
of $N$-representable pairs $\left(n,\mathbf{m}\right)$, characterized
in Ref.~\citep{Gontier2013}; it requires in particular that the
spin-density matrix be positive semidefinite, $|\mathbf{m}(\mathbf{r})|\le n(\mathbf{r})$.

\emph{Global spin-rotation symmetry.} As discussed above, in the Hamiltonian
Eq.~\ref{eq:Many-body-hamiltonian} the spin couples only to the
external field $\mathbf{B}_{\mathrm{ext}}$ and its spin structure
is therefore constrained by the behavior under a global rotation of
all spins. We will describe a spin rotation using 3D rotation matrices
in real space as well as the corresponding unitary operators in Hilbert
space. Let $\mathcal{R}_{\bm{\omega}}$ be a constant rotation by
an angle $\omega$ about the axis $\hat{\mathbf{n}}$, with rotation
vector $\bm{\omega}=\omega\hat{\mathbf{n}}$, and let 
\[
\hat{U}_{\bm{\omega}}=e^{-i\bm{\omega}\cdot\hat{\mathbf{S}}}\,,\qquad\hat{\mathbf{S}}=\tfrac{1}{2}\sum_{i}\bm{\sigma}_{i},
\]
be the unitary describing the corresponding global spin rotation in
Hilbert space. The two operators are connected by their action on
the Pauli matrices (see Appendix~\ref{sec:Orientation-relaxed-self-consist})
\begin{equation}
\hat{U}^{\dagger}_{\bm{\omega}}\bm{\sigma}_{i}\hat{U}_{\bm{\omega}}=\mathcal{R}_{\bm{\omega}}\bm{\sigma}_{i},\label{eq:Pauli-sigma-under-spin-rotation}
\end{equation}
so that under $\Psi\to\hat{U}_{\bm{\omega}}\Psi$ the charge density
is unchanged while the magnetization rotates uniformly at each point,
\begin{equation}
n(\mathbf{r})\to n(\mathbf{r}),\qquad\mathbf{m}(\mathbf{r})\to\mathcal{R}_{\bm{\omega}}\mathbf{m}(\mathbf{r}).\label{eq:magnetization-under-spin-rotation}
\end{equation}

Since $\hat{T}$ and $\hat{W}$ commute with $\hat{\mathbf{S}}$,
they are invariant under $\hat{U}_{\bm{\omega}}$, while the external
field must transform to preserve the scalar products $\mathbf{B}_{\mathrm{ext}}\cdot\bm{\sigma}_{i}$
in the Hamiltonian in accordance with Eq.~(\ref{eq:Pauli-sigma-under-spin-rotation}):
\begin{equation}
\hat{U}^{\dagger}_{\bm{\omega}}\hat{H}[v_{\mathrm{ext}},\mathbf{B}_{\mathrm{ext}}]\hat{U}_{\bm{\omega}}=\hat{H}[v_{\mathrm{ext}},\mathcal{R}^{\mathsf{T}}_{\bm{\omega}}\mathbf{B}_{\mathrm{ext}}].\label{eq:H-unitary-equivalence-to-spin-rotation}
\end{equation}
Rotating the field therefore produces a unitarily equivalent Hamiltonian,
with the same spectrum and with eigenstates rotated by $\hat{U}_{\bm{\omega}}$.
The energy can depend on $\mathbf{B}_{\mathrm{ext}}$ only through
rotational invariants such as $\mathbf{B}_{\mathrm{ext}}(\mathbf{r})\cdot\mathbf{B}_{\mathrm{ext}}(\mathbf{r}')$;
for uniform external fields, which are usually those of interest,
only the magnitude $B_{\mathrm{ext}}$ matters. The same holds for
the densities: physical predictions depend only on $n(\mathbf{r})$
and on the rotational invariants built from $\mathbf{m}$ and $\mathbf{B}_{\mathrm{ext}}$
together, e.g., pair products $\mathbf{m}(\mathbf{r})\cdot\mathbf{m}(\mathbf{r}')$
and $\mathbf{m}(\mathbf{r})\cdot\mathbf{B}_{\mathrm{ext}}(\mathbf{r}')$,
but not on their separate absolute direction. In particular the local
magnitude $|\mathbf{m}(\mathbf{r})|$, the angle between the magnetization
and the applied field, and the magnitude of the total moment $\mathbf{M}=\int\mathbf{m}(\mathbf{r})\,d^{3}r=2\langle\hat{\mathbf{S}}\rangle$
are all physical quantities, while the\emph{ direction }of $\mathbf{M}$
is physical only if $\mathbf{B}_{\mathrm{ext}}\ne0$, in which case
it must be antiparallel to the field.

At $\mathbf{B}_{\mathrm{ext}}=0$ the Hamiltonian commutes with $\hat{\mathbf{S}}$,
and $S$ is a good quantum number with a $(2S+1)$-fold degenerate
multiplet $|S,S^{z}\rangle$, connected by spin rotations and labeled
by $S^{z}=-S,\dots,S$. The basis states $|S,S^{z}\rangle$ have $|\mathbf{M}|=2|S^{z}|$.
Mixed states --- and, for $S\ge1$, suitable superpositions ---
can realize any $\lvert\mathbf{M}\rvert$ between $0$ and $2S$,
all with the same energy.

\emph{Spin density functional theory.} Following the Kohn--Sham approach\citep{KohnSham1965,vonBarthHedin1972},
we write 
\begin{equation}
F[n,\mathbf{m}]=T_{\mathrm{s}}[n,\mathbf{m}]+E_{\mathrm{H}}[n]+E_{\mathrm{xc}}[n,\mathbf{m}],\label{eq:F-in-Kohn-Sham-approach}
\end{equation}
with $T_{\mathrm{s}}[n,\mathbf{m}]=\min_{\Gamma_{\mathrm{s}}\to(n,\mathbf{m})}\mathrm{Tr}\bigl[\hat{T}\Gamma_{\mathrm{s}}\bigr]$
(the minimum now restricted to non-interacting $N$-electron density
matrices) and $E_{\mathrm{H}}[n]=\frac{1}{2}\int\frac{n(\mathbf{r})n(\mathbf{r}')}{|\mathbf{r}-\mathbf{r}'|}\,d^{3}r\,d^{3}r'$.
This defines $E_{\mathrm{xc}}[n,\mathbf{m}]$, the central object
for which we must make approximations.

Because $\hat{U}_{\bm{\omega}}$ commutes with $\hat{T}+\hat{W}$
and, by Eq.~(\ref{eq:magnetization-under-spin-rotation}), maps the
wavefunctions yielding $(n,\mathbf{m})$ one to one onto those yielding
$(n,\mathcal{R}_{\bm{\omega}}\mathbf{m})$, the two constrained searches
in Eq.~(\ref{eq:constrained-search}) have the same minimum. Hence
\begin{equation}
E_{\mathrm{xc}}[n,\mathcal{R}_{\bm{\omega}}\mathbf{m}]=E_{\mathrm{xc}}[n,\mathbf{m}]\qquad\text{for every constant }\bm{\omega}.\label{eq:Exc-invariance-To-Spin-Rotation}
\end{equation}
Equation~(\ref{eq:Exc-invariance-To-Spin-Rotation}) is an exact
condition on the functional. This is a strong statement, and since
$E_{\mathrm{xc}}$ is independent of the external potentials, it holds
for every system, independent of size, boundary conditions, and the
magnitude of the external fields. 

We use the Kohn--Sham approach to replace the interacting problem
by non-interacting spinors $\varphi_{i}$, eigenstates of the KS Hamiltonian,
\begin{equation}
\Bigl[-\tfrac{1}{2}\nabla^{2}+v_{s}(\mathbf{r})+\bm{\sigma}\cdot\mathbf{B}_{s}(\mathbf{r})\Bigr]\varphi_{i}=\varepsilon_{i}\varphi_{i},\label{eq:Kohn-Sham-equations}
\end{equation}
reproducing the densities $n=\sum_{i}\varphi^{\dagger}_{i}\varphi_{i}$
and $\mathbf{m}=\sum_{i}\varphi^{\dagger}_{i}\bm{\sigma}\varphi_{i}$.
The effective potentials are 
\[
v_{s}=v_{\mathrm{ext}}+v_{\mathrm{H}}+v_{\mathrm{xc}},\qquad\mathbf{B}_{s}=\mathbf{B}_{\mathrm{ext}}+\mathbf{B}_{\mathrm{xc}},
\]
where $v_{\mathrm{H}}(\mathbf{r})=\int n(\mathbf{r}')/|\mathbf{r}-\mathbf{r}'|\,d^{3}r'$
is the Hartree potential, the classical electrostatic potential of
the charge density, and the exchange-correlation potentials are defined
by the first variation of $E_{\mathrm{xc}}$, 
\begin{equation}
\delta E_{\mathrm{xc}}=\int v_{\mathrm{xc}}(\mathbf{r})\,\delta n(\mathbf{r})\,d^{3}r+\int\mathbf{B}_{\mathrm{xc}}(\mathbf{r})\cdot\delta\mathbf{m}(\mathbf{r})\,d^{3}r.\label{eq:vxc_Bxc_def}
\end{equation}
Note that $\mathbf{B}_{\mathrm{xc}}$ is defined here as a functional
derivative: we do not regard $\mathbf{B}_{\mathrm{xc}}$ as a physical
magnetic field.

\emph{Spin torques. }Next, consider the effect of spin rotation on
the energy. We write the derivative of the total energy, Eq.~(\ref{eq:total-energy-functional}),
with respect to a rigid rotation of the magnetization about an axis
$\hat{\mathbf{n}}$ as
\begin{equation}
\partial_{\hat{\mathbf{n}}}E\equiv\left.\frac{d}{d\omega}E\big[n,\mathcal{R}_{\omega\hat{\mathbf{n}}}\mathbf{m}\big]\right|_{\omega=0}.
\end{equation}
A global spin rotation leaves the density untouched, $\delta n=0$,
and changes the magnetization by $\delta\mathbf{m}=\omega\,\hat{\mathbf{n}}\times\mathbf{m}$
{[}Eq.~(\ref{eq:infinitesimal-rotation}){]}. Inserting these into
Eq.~(\ref{eq:vxc_Bxc_def}) gives $\partial_{\hat{\mathbf{n}}}E=\int\left(\mathbf{B}_{\mathrm{ext}}(\mathbf{r})+\mathbf{B}_{\mathrm{xc}}(\mathbf{r})\right)\cdot\hat{\mathbf{n}}\times\mathbf{m}\,d^{3}r$
which evaluates to
\begin{equation}
\partial_{\hat{\mathbf{n}}}E=\hat{\mathbf{n}}\cdot\left(\bm{\mathcal{T}}_{\mathrm{ext}}+\bm{\mathcal{T}}_{\mathrm{xc}}\right),\label{eq:energy-change}
\end{equation}
where $\bm{\mathcal{T}}_{\mathrm{ext}}=\int\mathbf{m}\times\mathbf{B}_{\mathrm{ext}}(\mathbf{r})d^{3}r$
is the external field torque, the derivative of the Zeeman energy
$\int\mathbf{B}_{\mathrm{ext}}\cdot\mathbf{m}\,d^{3}r$. The second
term $\bm{\mathcal{T}}_{\mathrm{xc}}=\int\mathbf{m}\times\mathbf{B}_{\mathrm{xc}}(\mathbf{r})d^{3}r$
is the xc torque describing the change in xc energy:
\begin{equation}
\partial_{\hat{\mathbf{n}}}E_{\mathrm{xc}}=\hat{\mathbf{n}}\cdot\int\bm{\tau}_{\mathrm{xc}}\left(\mathbf{r}\right)\,d^{3}r,\label{eq:torque-identity}
\end{equation}
where $\bm{\tau}_{\mathrm{xc}}=\mathbf{m}\times\mathbf{B}_{\mathrm{xc}}$
is the local xc torque. If $E_{\mathrm{xc}}$ obeys Eq.~(\ref{eq:Exc-invariance-To-Spin-Rotation})
the left side of Eq.~(\ref{eq:torque-identity}) vanishes for every
$\hat{\mathbf{n}}$, and with it the total xc torque, 
\begin{equation}
\bm{\mathcal{T}}_{\mathrm{xc}}\equiv\int\bm{\tau}_{\mathrm{xc}}(\mathbf{r})\,d^{3}r=0.\label{eq:zero-xc-torque-condition}
\end{equation}
This is the ``zero xc spin-torque theorem'', first derived in \citep{CapelleVignaleGyorffy2001}.
It reflects the fact that the electron--electron interaction cannot
rotate the total spin on its own. Conversely, when a functional violates
Eq.~(\ref{eq:Exc-invariance-To-Spin-Rotation}) the total torque
no longer vanishes: its component along any axis is the angular gradient
of $E_{\mathrm{xc}}$ about that axis.

Returning to Eq.~(\ref{eq:energy-change}), at a self-consistent
solution of the KS equations the energy is stationary against every
allowed variation of the densities, a rigid spin rotation among them,
so the left-hand side vanishes, giving zero total torque:
\begin{equation}
\bm{\mathcal{T}}_{\mathrm{ext}}+\bm{\mathcal{T}}_{\mathrm{xc}}=0.\label{eq:zero-total-torque-condition}
\end{equation}
For a functional obeying Eq.~(\ref{eq:Exc-invariance-To-Spin-Rotation}),
the xc term vanishes separately by Eq.~(\ref{eq:zero-xc-torque-condition}),
and thus 
\begin{equation}
\bm{\mathcal{T}}_{\mathrm{ext}}\equiv\int\mathbf{m}\times\mathbf{B}_{\mathrm{ext}}d^{3}r=0.\label{eq:zero-ext-torque-condition}
\end{equation}
For a uniform field this translates to $\mathbf{M}\times\mathbf{B}_{\mathrm{ext}}=0$,
showing that $\mathbf{M}$ must be collinear with $\mathbf{B}_{\mathrm{ext}}$
(or vanish). Of the two collinear orientations the lower in energy
is the antiparallel one.

\emph{Locally collinear approximations.} Below we follow the common
practice of most non-collinear calculations and take a functional
developed for collinear systems and evaluate it on the magnitude $m=|\mathbf{m}|$
\citep{Kubler1988,PeraltaScuseriaFrisch2007}. For the local spin
density approximation (LSDA), $E^{\mathrm{LSDA}}_{\mathrm{xc}}[n,\mathbf{m}]=\int\varepsilon^{\mathrm{LSDA}}_{\mathrm{xc}}(n(\mathbf{r}),m(\mathbf{r}))\,d^{3}r$,
Eq.~(\ref{eq:vxc_Bxc_def}) gives 
\begin{equation}
\mathbf{B}^{\mathrm{LSDA}}_{\mathrm{xc}}(\mathbf{r})=\varepsilon^{\mathrm{LSDA}}_{\mathrm{xc},m}(n,m)\,\frac{\mathbf{m}(\mathbf{r})}{m(\mathbf{r})},\qquad\varepsilon^{\mathrm{LSDA}}_{\mathrm{xc},m}\equiv\frac{\partial\varepsilon^{\mathrm{LSDA}}_{\mathrm{xc}}}{\partial m},\label{eq:LSDA-field}
\end{equation}
a field everywhere parallel to the local magnetization. Eq.~(\ref{eq:LSDA-field})
has an important consequence: the local exchange-correlation torque
density of such functionals vanishes identically, $\mathbf{m}\times\mathbf{B}^{\mathrm{LSDA}}_{\mathrm{xc}}=0$.
This lack of local torques is a drawback of the local exchange correlation
functionals, as they lack a direct mechanism for the exchange-correlation
energy to reorient the magnetization. There is ongoing work on functionals
with nonzero local torques \citep{ScalmaniFrisch2012,Bulik2013,EichGross2013,Pu2023,PluharUllrich2019,TancogneDejeanRubioUllrich2023,LiraPeralta2026}.

\emph{Source-free exchange-correlation fields.} It has been argued
\citep{CapelleGross1997,Sharma2018} that the exchange-correlation
field should be source-free, 
\begin{equation}
\nabla\cdot\mathbf{B}_{\mathrm{xc}}=0,\label{eq:source-free-condition}
\end{equation}
like a physical magnetic field. Sharma and coworkers impose Eq.~(\ref{eq:source-free-condition})
by projecting the longitudinal (source) component out of the field
a posteriori, and report clearly improved magnetic moments and non-collinear
ground states in many materials \citep{Sharma2018,Krishna2019,Dewhurst2018}.
The idea has since been taken up independently, with implementations
in the CASTEP \citep{Hawkhead2026} and VASP \citep{Moore2025} plane-wave
codes reporting similar improvements for frustrated magnets. The projected
field, however, is not the derivative of an energy functional. The
Kohn--Sham problem is then no longer variational, the exchange-correlation
kernel loses the symmetry of a second derivative: forces and response
properties cease to be consistent with the derivatives of the energy,
in particular, the Hellmann--Feynman theorem does not apply. The
zero xc spin-torque theorem, Eq.~(\ref{eq:zero-xc-torque-condition}),
fails as well. Ref.~\citep{Moore2025} shows that for a source-free
field $\mathbf{B}^{\mathrm{SF}}_{\mathrm{xc}}=\mathbf{B}^{\mathrm{parent}}_{\mathrm{xc}}+\nabla\chi$
built on a locally collinear parent the net torque is $\bm{\mathcal{T}}_{\mathrm{xc}}=\int\chi\left(\nabla\times\mathbf{m}\right)d^{3}r$,
which vanishes only for a curl-free magnetization. They find it small
in practice, of the order of their energy convergence, and propose
a way to impose it. In the dynamical regime, ref.~\citep{BologaUllrich2024}
found large errors that they traced to this non-variationality: Larmor
frequencies are off by a large percentage, and the ferromagnetic magnon
dispersion is linear in wavevector instead of quadratic. A variational
source-free field was still missing.

\emph{This work.} Our original motivation was to present an ansatz
that turns any parent exchange-correlation functional into one whose
xc field is source-free, Eq.~(\ref{eq:source-free-condition}), and
remains a functional derivative. The new functional also reduces to
its parent for the homogeneous electron gas, which we regarded as
a strong point. We implemented it in PySCF, and the Kohn--Sham equations
indeed produced a source-free $\mathbf{B}_{\mathrm{xc}}$ that was
variational and obeyed the zero xc spin-torque condition at self-consistency.
But problems soon appeared: in an applied field the magnetization
pointed in the wrong direction. We traced this to the new $E_{\mathrm{xc}}$
breaking global spin-rotation symmetry, and then further, to the source-free
condition itself, which no invariant functional can satisfy.

Our work thus evolved. In Sec.~\ref{sec:impossibility} we show that
a source-free field for the $E_{\mathrm{xc}}$ of SDFT cannot be an
exact condition. In Sec.~\ref{sec:transverse-theory} we work out
a transverse theory, $\perp$SDFT, following Ref.~\citep{Sharma2018},
in which a source-free field does exist and which gives the same energy,
density and transverse magnetization as SDFT. But its exchange-correlation
energy differs from that of SDFT, and its Kohn--Sham system does
not deliver the longitudinal magnetization. The benefit of $\perp$SDFT
could be in finding new approximations. Sec.~\ref{sec:The-SoF-construction}
presents ours, the bare-parent approximation, which is the ansatz
we started from. Sec.~\ref{sec:manganese-dimer} takes \ce{Mn2}
as a case study and examines the consequences of using it: the magnetic
improvement reported with source-free fields and, beside it, the deleterious
effects of the broken symmetry and their magnitude.

\section{Source-freeness cannot be an exact condition of SDFT}\label{sec:impossibility}

The global spin-rotation symmetry forbids the claim that $\mathbf{B}_{\mathrm{xc}}$
is source-free. To see why, we first examine how $\mathbf{B}_{\mathrm{xc}}$
transforms under a global spin rotation. Using Eq.~(\ref{eq:vxc_Bxc_def})
while noting that spin rotations do not affect the density $n$,\emph{
}we differentiate Eq.~(\ref{eq:Exc-invariance-To-Spin-Rotation})
with respect to the magnetization; the chain rule then yields $\mathcal{R}^{\mathsf{T}}_{\bm{\omega}}\mathbf{B}_{\mathrm{xc}}[n,\mathcal{R}_{\bm{\omega}}\mathbf{m}]=\mathbf{B}_{\mathrm{xc}}[n,\mathbf{m}]$,
and since for rotation matrices $\mathcal{R}^{\mathsf{T}}_{\bm{\omega}}=\mathcal{R}^{-1}_{\bm{\omega}}$
the transformation law of the field follows: 
\begin{equation}
\mathbf{B}_{\mathrm{xc}}[n,\mathcal{R}_{\bm{\omega}}\mathbf{m}]=\mathcal{R}_{\bm{\omega}}\mathbf{B}_{\mathrm{xc}}[n,\mathbf{m}].\label{eq:covariance-Bxc-to-Spin-Rotation}
\end{equation}
We see that the field may depend on the direction of $\mathbf{m}$,
but when the magnetization is rigidly rotated the resulting xc field
rotates with it. The simplest example for this is $\mathbf{B}^{\mathrm{LSDA}}_{\mathrm{xc}}$
of Eq.~(\ref{eq:LSDA-field}), which is simply parallel to $\mathbf{m}$. 

Suppose now that Eq.~(\ref{eq:source-free-condition}) were an exact
condition, holding for every admissible magnetization. By Eq.~(\ref{eq:Exc-invariance-To-Spin-Rotation})
a rotated magnetization $\mathcal{R}_{\bm{\omega}}\mathbf{m}$ leaves
$E_{\mathrm{xc}}$ unchanged, and by Eq.~(\ref{eq:covariance-Bxc-to-Spin-Rotation})
the field belonging to it is $\mathcal{R}_{\bm{\omega}}\,\mathbf{B}_{\mathrm{xc}}$.
We would then have to demand, by hypothesis, $\nabla\cdot(\mathcal{R}_{\bm{\omega}}\,\mathbf{B}_{\mathrm{xc}})=0$
for every $\bm{\omega}=\omega\hat{\mathbf{n}}$. But by Eq.~(\ref{eq:div-rotation})
of Appendix~\ref{sec:TheTransverseProjection} a global rotation
mixes a field's curl into its divergence through the term $\hat{\mathbf{n}}\cdot(\nabla\times\mathbf{B}_{\mathrm{xc}})$,
which must therefore vanish as well. The axis $\hat{\mathbf{n}}$
is arbitrary, so $\nabla\times\mathbf{B}_{\mathrm{xc}}=0$. The field
would then have neither sources nor a curl, which means it is either
uniform or zero --- too trivial to represent exchange and correlation
in an inhomogeneous system. We conclude that Eq.~(\ref{eq:source-free-condition})
is forbidden by the symmetry expressed in Eq.~(\ref{eq:Exc-invariance-To-Spin-Rotation}):
it is not a valid condition.

The argument rests on two properties of $E_{\mathrm{xc}}$: its invariance
under global spin rotations, Eq.~(\ref{eq:Exc-invariance-To-Spin-Rotation}),
which is exact, and its differentiability with respect to $\mathbf{m}$
at the magnetizations considered, so that $\mathbf{B}_{\mathrm{xc}}$
exists and obeys Eq.~(\ref{eq:covariance-Bxc-to-Spin-Rotation}).
The second property is also the premise of the source-free condition
itself: without a functional derivative there is no field whose divergence
could vanish. Our conclusion is therefore that wherever $\mathbf{B}_{\mathrm{xc}}$
is defined it cannot be source-free for every admissible $\mathbf{m}$.

This bears on current practice. Modern non-collinear functionals are
invariant under global spin rotations, and this invariance coexists
with nonvanishing local xc torques: examples are the Scalmani--Frisch
formalism\citep{ScalmaniFrisch2012,Bulik2013} and its current-density
extension\citep{Franzke2025}, the multicollinear approach\citep{Pu2023,Wang2026},
which for an LDA parent reduces to the locally collinear form we use
below, and semilocal functionals built to carry xc torques\citep{TancogneDejeanRubioUllrich2023}.
The meta-GGAs of spin-current DFT go further and are invariant under
local $U(1)\times SU(2)$ gauge transformations\citep{PittalisVignaleEich2017,Desmarais2024},
of which global spin rotations are the constant subgroup. All of these
functionals obey Eq.~(\ref{eq:Exc-invariance-To-Spin-Rotation}),
so by the argument above none can have a source-free $\mathbf{B}_{\mathrm{xc}}$,
and none is claimed to. The same conclusion is reached independently
within spin-current density functional theory, whose basic variables
are the particle and spin densities together with the paramagnetic
charge and spin currents \citep{PittalisVignaleEich2017}. There the
xc scalar potential, vector potential and magnetic field are shown
to be effective Yang--Mills rather than Maxwellian fields, with $\mathbf{B}_{\mathrm{xc}}\neq\nabla\times\mathbf{A}_{\mathrm{xc}}$
in general, so that $\mathbf{B}_{\mathrm{xc}}$ inherits no source-free
property from a vector potential. Because $E_{\mathrm{xc}}$ is universal
and does not refer to any specific external magnetic fields the presence
or absence of $\mathbf{B}_{\mathrm{ext}}$ does not change the conclusions. 

A source-free field can nevertheless be obtained by restricting the
theory to the transverse part of the magnetization, as Sharma et al.~\citep{Sharma2018}
do. The next section shows what that restriction amounts to.

\section{$\perp$SDFT: Transverse spin density functional theory}\label{sec:transverse-theory}

Sharma et al.~\citep{Sharma2018} obtain a source-free field by a
different route. If the external magnetic field is physical, $\mathbf{B}_{{\rm ext}}=\nabla\times\mathbf{A}_{{\rm ext}}$,
it couples only to $\nabla\times\mathbf{m}$, and they define by constrained
search at fixed $(n,\nabla\times\mathbf{m})$ a universal functional
$\tilde{F}$, a non-interacting kinetic energy $\tilde{T}_{{\rm s}}$,
and an exchange-correlation energy $\tilde{E}_{{\rm xc}}$, whose
field is source-free by construction. They state, after Eq.~(8) of
Ref.~\citep{Sharma2018}, that on the densities of physical external
fields $\tilde{E}_{{\rm xc}}$ coincides with $E_{{\rm xc}}$, and
note that the Kohn--Sham magnetization of this theory is exact only
in its curl and, for finite systems, in its total moment. We now work
this theory out with the transverse magnetization $\mathbf{m}_{\perp}$
as the variable, and call it transverse spin-density functional theory
($\perp$SDFT); $\mathbf{m}_{\perp}$ carries the same information
as $\nabla\times\mathbf{m}$ together with the total moment that Ref.~\citep{Sharma2018}
adds for periodic systems (Appendix~\ref{sec:TheTransverseProjection}).
However, the two variables differ in locality: $\nabla\times\mathbf{m}$
vanishes wherever $\mathbf{m}$ does, whereas $\mathbf{m}_{\perp}$,
its Biot--Savart transform, extends beyond the system with the algebraic
tail of Eq.~(\ref{eq:dipole-field}); a functional local in one is
nonlocal in the other. This may affect approximations (Sec.~\ref{sec:The-SoF-construction})
but not the exact theory, which we now develop. 

\emph{The interacting theory.} Consider the Hohenberg--Kohn functional
\begin{equation}
E[v,\mathbf{B}_{{\rm ext}}]=\min_{n,\mathbf{m}}\Big\{ F[n,\mathbf{m}]+\int v\,n+\int\mathbf{B}_{{\rm ext}}\cdot\mathbf{m}\Big\}.\label{eq:E_HK}
\end{equation}
Writing $\mathbf{m}=\mathbf{m}_{\perp}+\nabla\chi$, Eq.~(\ref{eq:E_HK})
is trivially equivalent to \begin{widetext}
\begin{equation}
E[v,\mathbf{B}_{{\rm ext}}]=\min_{n,\mathbf{m}_{\perp}}\min_{\chi}\Big\{ F[n,\mathbf{m}_{\perp}+\nabla\chi]+\int v\,n+\int\mathbf{B}_{{\rm ext}}\cdot(\mathbf{m}_{\perp}+\nabla\chi)\Big\}.
\end{equation}
\end{widetext}If $\nabla\cdot\mathbf{B}_{{\rm ext}}=0$ then $\int\mathbf{B}_{{\rm ext}}\cdot\nabla\chi=0$,
the inner minimization applies only to $F$, and 
\begin{equation}
E[v,\mathbf{B}_{{\rm ext}}]=\min_{n,\mathbf{m}_{\perp}}\Big\{\tilde{F}[n,\mathbf{m}_{\perp}]+\int v\,n+\int\mathbf{B}_{{\rm ext}}\cdot\mathbf{m}_{\perp}\Big\},
\end{equation}
where
\begin{equation}
\tilde{F}[n,\mathbf{m}]=\min_{\chi}F[n,\mathbf{m}+\nabla\chi].\label{eq:F-tilde}
\end{equation}
We define $\tilde{F}$ for every magnetization, not only for transverse
ones: since $\mathbf{m}+\nabla\chi$ runs over the same set for all
$\mathbf{m}$ with the same transverse part, $\tilde{F}[n,\mathbf{m}]=\tilde{F}[n,\mathbf{m}_{\perp}]$,
i.e.~$\tilde{F}$ is blind to the longitudinal part of its argument.
Let $\chi_{*}[n,\mathbf{m}_{\perp}]$ denote the inner minimizer.
The outer minimum is attained at $n_{*}$ and $(\mathbf{m}_{*})_{\perp}$
(where $n_{*}$ and $\mathbf{m}_{*}$ are densities of the ground
state). The full magnetization is recovered from the inner minimizer,
$\mathbf{m}_{*}=(\mathbf{m}_{*})_{\perp}+\nabla\chi_{*}[n_{*},(\mathbf{m}_{*})_{\perp}]$.
At this level nothing is lost: for source-free external fields the
interacting theory can be formulated with $\mathbf{m}_{\perp}$ alone
as the magnetic variable and $\tilde{F}$ as the universal functional,
and it gives the exact energy, density and magnetization.

\emph{The Kohn--Sham theory.} Take any exchange-correlation functional
of SDFT, exact or approximate, and let $F=T_{{\rm s}}+E_{{\rm H}}+E_{{\rm xc}}$
be the universal functional built on it, so that Eqs.~(\ref{eq:E_HK})--(\ref{eq:F-tilde})
hold with $E[v,\mathbf{B}_{{\rm ext}}]$, $n_{*}$ and $\mathbf{m}_{*}$
the ground-state energy and densities of that theory. Define the non-interacting
kinetic energy of $\perp$SDFT as $\tilde{F}$ was defined, 
\begin{equation}
\tilde{T}_{{\rm s}}[n,\mathbf{m}]=\min_{\chi}T_{{\rm s}}[n,\mathbf{m}+\nabla\chi],\label{eq:tilde-Ts}
\end{equation}
and its exchange-correlation energy as the remainder, $\tilde{E}_{{\rm xc}}=\tilde{F}-E_{{\rm H}}-\tilde{T}_{{\rm s}}$.
Since $E_{{\rm H}}$ depends on $n$ alone, $\tilde{F}-E_{{\rm H}}=\min_{\chi}(T_{{\rm s}}+E_{{\rm xc}})[n,\mathbf{m}+\nabla\chi]$,
and \begin{widetext}
\begin{equation}
\tilde{E}_{{\rm xc}}[n,\mathbf{m}]=\min_{\chi}\left\{ \left(T_{{\rm s}}+E_{{\rm xc}}\right)[n,\mathbf{m}+\nabla\chi]\right\} -\min_{\chi}T_{{\rm s}}[n,\mathbf{m}+\nabla\chi].\label{eq:Exc-tilde-explicit}
\end{equation}
Like $\tilde{F}$, both $\tilde{T}_{{\rm s}}$ and $\tilde{E}_{{\rm xc}}$
are defined for every $\mathbf{m}$ and blind to its longitudinal
part. The Kohn--Sham energy of $\perp$SDFT is then the Kohn--Sham
energy of SDFT minimized over the longitudinal magnetization, 
\begin{equation}
\left(\tilde{T}_{{\rm s}}+E_{{\rm H}}+\tilde{E}_{{\rm xc}}\right)[n,\mathbf{m}]=\min_{\chi}\left(T_{{\rm s}}+E_{{\rm H}}+E_{{\rm xc}}\right)[n,\mathbf{m}+\nabla\chi],\label{eq:re-partitioning}
\end{equation}
and minimizing it over $(n,\mathbf{m}_{\perp})$ returns $E[v,\mathbf{B}_{{\rm ext}}]$,
$n_{*}$ and $(\mathbf{m}_{*})_{\perp}$. 

\emph{The Kohn--Sham system.} Kohn--Sham orbitals produce a full
magnetization $\mathbf{m}[\Phi]=\sum_{i}\varphi^{\dagger}_{i}\bm{\sigma}\varphi_{i}$,
longitudinal part included, so the orbital energy functional reads
\begin{equation}
E[\Phi]=T_{{\rm s}}[\Phi]+E_{{\rm H}}[n]+\tilde{E}_{{\rm xc}}\big[n,\mathbf{m}[\Phi]\big]+\int v\,n+\int\mathbf{B}_{{\rm ext}}\cdot\mathbf{m}[\Phi],\label{eq:KS-orbital-functional}
\end{equation}
\end{widetext}where the Zeeman term needs no projection, since $\int\mathbf{B}_{{\rm ext}}\cdot\mathbf{m}=\int\mathbf{B}_{{\rm ext}}\cdot\mathbf{m}_{\perp}$
for a source-free field, and minimizing $T_{{\rm s}}[\Phi]$ at fixed
$n$ and $\mathbf{m}_{\perp}$ realizes $\tilde{T}_{{\rm s}}$. In
Eq.~(\ref{eq:KS-orbital-functional}) $\tilde{E}_{{\rm xc}}$ is
a functional of the full magnetization, blind to its longitudinal
part; conversely, every functional of $\mathbf{m}$ invariant under
$\mathbf{m}\to\mathbf{m}+\nabla\chi$ is of this kind. So $\perp$SDFT
is SDFT with $E_{{\rm xc}}$ replaced by a longitudinally blind functional.\footnote{Note, that it is in this form that we approximate it in Sec.~\ref{sec:The-SoF-construction}.}
Varying the orbitals varies $\mathbf{m}$ freely, and the field entering
the Kohn--Sham equations is the functional derivative
\begin{equation}
\tilde{\mathbf{B}}_{{\rm xc}}=\frac{\delta\tilde{E}_{{\rm xc}}}{\delta\mathbf{m}}.\label{eq:tildeExc-deriv}
\end{equation}
Because $\tilde{E}_{{\rm xc}}$ is blind to gradients, $\langle\tilde{\mathbf{B}}_{{\rm xc}},\nabla\chi\rangle=0$
for every $\chi$, i.e.~$\nabla\cdot\tilde{\mathbf{B}}_{{\rm xc}}=0$:
the field is transverse, and so is $\tilde{\mathbf{B}}_{s}=\mathbf{B}_{{\rm ext}}+\tilde{\mathbf{B}}_{{\rm xc}}$.
This is the source-free field. It is the derivative of $\tilde{E}_{{\rm xc}}$,
not of $E_{{\rm xc}}$, so there is no conflict with Sec.~\ref{sec:impossibility}.

\emph{What is lost.} The Kohn--Sham determinant reproduces $n_{*}$
and $(\mathbf{m}_{*})_{\perp}$ but not $\mathbf{m}_{*}$: its longitudinal
magnetization is the minimizer of $T_{{\rm s}}$ in Eq.~(\ref{eq:tilde-Ts}),
not $\nabla\chi_{*}$. The longitudinal part of $\mathbf{m}_{*}$
is recoverable in principle from the inner minimization defining $\tilde{F}$,
but not from the orbitals. Local moments, canting angles and spin
textures, which depend on $\mathbf{m}$ pointwise, are therefore inaccessible
to the Kohn--Sham system of $\perp$SDFT. These are the very quantities
for which source-free fields are credited with improvements \citep{Sharma2018,Krishna2019,Dewhurst2018,Hawkhead2026,Moore2025}.
Exact $\perp$SDFT does not deliver them at all, so the improvements
cannot come from the source-free condition itself; they may come from
the approximations that impose it. The total moment $\mathbf{M}_{*}=\int(\mathbf{m}_{*})_{\perp}\,d^{3}r$
remains physical \citep{Sharma2018}, the longitudinal part integrating
to zero for a finite system. 

\emph{The exchange-correlation energy is not that of SDFT.} At the
ground state the first minimization in Eq.~(\ref{eq:Exc-tilde-explicit})
is attained at $\mathbf{m}_{*}$, so 
\[
\left(\tilde{T}_{{\rm s}}+\tilde{E}_{{\rm xc}}\right)\left[n_{*},\mathbf{m}_{*}\right]=\left(T_{{\rm s}}+E_{{\rm xc}}\right)\left[n_{*},\mathbf{m}_{*}\right],
\]
while by definition $\tilde{T}_{{\rm s}}[n_{*},\mathbf{m}_{*}]\le T_{{\rm s}}[n_{*},\mathbf{m}_{*}]$.
Hence\begin{widetext} 
\begin{equation}
\tilde{E}_{{\rm xc}}\left[n_{*},\mathbf{m}_{*}\right]-E_{{\rm xc}}\left[n_{*},\mathbf{m}_{*}\right]=T_{{\rm s}}[n_{*},\mathbf{m}_{*}]-\tilde{T}_{{\rm s}}[n_{*},\mathbf{m}_{*}]\ge0,\label{eq:Exc-gap}
\end{equation}
\end{widetext}with equality only if $\mathbf{m}_{*}$ itself minimizes
$T_{{\rm s}}$ over longitudinal variations, i.e., only if the Kohn--Sham
field $\mathbf{B}_{s}$ of SDFT is already source-free.\footnote{Stationarity of $T_{{\rm s}}[n_{*},\mathbf{m}_{*}+\epsilon\nabla\chi]$
at $\epsilon=0$ for every $\chi$ gives $\nabla\cdot(\delta T_{{\rm s}}/\delta\mathbf{m})=0$
at $\mathbf{m}_{*}$, and the SDFT Kohn--Sham condition $\delta T_{{\rm s}}/\delta\mathbf{m}=-\mathbf{B}_{s}$
turns this into $\nabla\cdot\mathbf{B}_{s}=0$.} By Sec.~\ref{sec:impossibility} no invariant functional has a source-free
field for every magnetization, so equality in Eq.~(\ref{eq:Exc-gap})
fails for some ground states. It holds for special ones, such as the
homogeneous electron gas, whose $\mathbf{B}_{\mathrm{xc}}$ is uniform.

Ref.~\citep{Sharma2018} defines $\tilde{F}$, $\tilde{T}_{{\rm s}}$
and $\tilde{E}_{{\rm xc}}$ exactly as here {[}Eqs.~(6) and (8) there
and the sentence following{]} and states that, on the densities of
physical external fields, $E_{{\rm xc}}[n,\mathbf{m}]=\tilde{E}_{{\rm xc}}[n,\nabla\times\mathbf{m}]$.
Eq.~(\ref{eq:Exc-gap}) shows this cannot hold in general: the gap
is nonzero whenever the Kohn--Sham magnetization differs from $\mathbf{m}_{*}$,
as Ref.~\citep{Sharma2018} itself notes it does. Their further statement
that the total energies of the two theories coincide is correct, Eq.~(\ref{eq:re-partitioning}),
but because $\tilde{T}_{{\rm s}}+\tilde{E}_{{\rm xc}}=T_{{\rm s}}+E_{{\rm xc}}$
at the ground state, not because the exchange-correlation energies
agree. Ref.~\citep{CapelleGross1997} reaches equality by a different
route, which we examine in Appendix~\ref{sec:cdft-route}.

\emph{Symmetry.} $\tilde{E}_{{\rm xc}}$ is blind to $\mathbf{m}_{\parallel}$,
so its field is source-free, and by Sec.~\ref{sec:impossibility}
no functional with a source-free field satisfies Eq.~(\ref{eq:Exc-invariance-To-Spin-Rotation}).
Hence $\tilde{E}_{{\rm xc}}[n,\mathcal{R}_{\bm{\omega}}\mathbf{m}]\neq\tilde{E}_{{\rm xc}}[n,\mathbf{m}]$
in general, and the same holds for $\tilde{F}$ and $\tilde{T}_{{\rm s}}$.\footnote{Ref.~\citep{Sharma2018} uses its functional of $\nabla\times\mathbf{m}$
in the same way, as a functional of $\mathbf{m}$ that is differentiated
with respect to $\mathbf{m}$ {[}Eq.~(9) there{]}; it is blind to
$\mathbf{m}_{\parallel}$ as well, and the same conclusion applies
to it.} The external field is restricted too: a rotated non-uniform field
$\mathcal{R}_{\bm{\omega}}\mathbf{B}_{{\rm ext}}$ is in general not
source-free, Eq.~(\ref{eq:div-rotation}), so $\mathbf{B}_{\mathrm{ext}}$
and its rotated version $\mathcal{R}_{\bm{\omega}}\mathbf{B}_{\mathrm{ext}}$
are both admissible in $\perp$SDFT only when $\mathbf{B}_{\mathrm{ext}}$
is uniform. In SDFT, $F$, $T_{\mathrm{s}}$ and $E_{\mathrm{xc}}$
are each invariant; in $\perp$SDFT none of them is. The energy of
$\perp$SDFT is invariant nonetheless, a property of the solution
rather than of the functionals, because the minimum over $(n,\mathbf{m}_{\perp},\chi)$
ranges over all magnetizations and is the invariant ground-state energy
$E[v,\mathbf{B}_{{\rm ext}}]$. Hence, in both SDFT and $\perp$SDFT,
for a uniform external field the energy depends only on $|\mathbf{B}_{{\rm ext}}|$:
there is no magnetic anisotropy, the total moment $\mathbf{M}_{*}=\partial E/\partial\mathbf{B}_{{\rm ext}}$
is antiparallel to the field, and by Eq.~(\ref{eq:zero-total-torque-condition})
the xc torque $\bm{\mathcal{T}}_{{\rm xc}}=-\mathbf{M}_{*}\times\mathbf{B}_{{\rm ext}}$
vanishes at the solution, so the zero xc spin-torque theorem survives
there. If we approximate $\tilde{E}_{\mathrm{xc}}$, the resulting
theory has no exact energy behind it, and its non-invariance will
appear as a magnetic anisotropy, a misaligned moment and a non-zero
xc torque. In Sec.~\ref{sec:manganese-dimer} we measure how severe
such a violation can be.

\section{The bare-parent approximation to $\perp$SDFT: a source-free construction}\label{sec:The-SoF-construction}

Given a parent exchange-correlation functional of SDFT, $E^{\mathrm{parent}}_{\mathrm{xc}}[n,\mathbf{m}]$,
we approximate the exchange-correlation energy of $\perp$SDFT by
\begin{equation}
\tilde{E}_{\mathrm{xc}}[n,\mathbf{m}]\approx E^{\mathrm{parent}}_{\mathrm{xc}}[n,\mathbf{m}_{\perp}],\label{eq:SoF-Exc-definition}
\end{equation}
the parent energy of a fictitious system constrained to $\nabla\cdot\mathbf{m}=0$.
We call this the bare-parent approximation: the exact $\tilde{E}_{\mathrm{xc}}$,
that of Eq.~(\ref{eq:Exc-tilde-explicit}), is the parent dressed
by a minimization over the longitudinal magnetization, and Eq.~(\ref{eq:SoF-Exc-definition})
uses it bare: no minimization, $\chi=0$. In figures and tables we
label it SoF, for source-free. Its field follows from Sec.~\ref{sec:transverse-theory}:
with $\mathbf{B}^{\mathrm{parent}}_{\mathrm{xc}}=\delta E^{\mathrm{parent}}_{\mathrm{xc}}/\delta\mathbf{m}$
the field of the parent, Eq.~(\ref{eq:symmetricHelmholtzProj}) gives
the field entering the Kohn--Sham equations as 
\begin{equation}
\tilde{\mathbf{B}}_{\mathrm{xc}}=\bigl(\mathbf{B}^{\mathrm{parent}}_{\mathrm{xc}}[n,\mathbf{m}_{\perp}]\bigr)_{\perp}.\label{eq:Bxc-general}
\end{equation}
The field is transverse, and it is the functional derivative of an
energy, so the exchange-correlation kernel retains the symmetry of
a second derivative, $f^{ab}_{\mathrm{xc}}(\mathbf{r},\mathbf{r}')=f^{ba}_{\mathrm{xc}}(\mathbf{r}',\mathbf{r})$,
where 
\[
f^{ab}_{\mathrm{xc}}(\mathbf{r},\mathbf{r}'):=\frac{\delta\tilde{B}^{a}_{\mathrm{xc}}(\mathbf{r})}{\delta m_{b}(\mathbf{r}')}.
\]
Eq.~(\ref{eq:Bxc-general}) also shows how the construction relates
to the a posteriori projection of Refs.~\citep{Sharma2018,Dewhurst2018}.
Their field is $\bigl(\mathbf{B}^{\mathrm{parent}}_{\mathrm{xc}}[n,\mathbf{m}]\bigr)_{\perp}$,
the parent field evaluated at the full magnetization and then projected,
with an empirical scaling $s\approx1.1$ of the magnetization that
they report to be essential; ours has no scaling and evaluates the
parent at the projected magnetization before projecting the field,
and that is what makes it a functional derivative. Since a uniform
magnetization is purely transverse (Appendix~\ref{sec:TheTransverseProjection}),
the construction reduces to the parent for the homogeneous electron
gas. If the parent field were already source-free for every density
the two fields would coincide and the construction would change nothing.\footnote{If $\nabla\cdot\mathbf{B}^{\mathrm{parent}}_{\mathrm{xc}}=0$ held
for every density then $E^{\mathrm{parent}}_{\mathrm{xc}}[n,\mathbf{m}_{\perp}+t\,\mathbf{m}_{\parallel}]$
would be independent of $t$: with $\mathbf{m}_{\parallel}=\nabla\chi$,
$dE^{\mathrm{parent}}_{\mathrm{xc}}/dt=\langle\mathbf{B}^{\mathrm{parent}}_{\mathrm{xc}},\nabla\chi\rangle=-\langle\nabla\cdot\mathbf{B}^{\mathrm{parent}}_{\mathrm{xc}},\chi\rangle=0$.
Then $E^{\mathrm{parent}}_{\mathrm{xc}}[n,\mathbf{m}_{\perp}]=E^{\mathrm{parent}}_{\mathrm{xc}}[n,\mathbf{m}]$.}

\emph{Two readings. }Eq.~(\ref{eq:SoF-Exc-definition}) is at once
an approximation to $\tilde{E}_{\mathrm{xc}}$ of $\perp$SDFT and,
inserted in Eq.~(\ref{eq:KS-orbital-functional}), an SDFT functional
$E_{\mathrm{xc}}[n,\mathbf{m}]:=E^{\mathrm{parent}}_{\mathrm{xc}}[n,\mathbf{m}_{\perp}]$
with a source-free field, which Sec.~\ref{sec:impossibility} holds
to the exact conditions of SDFT, Eq.~(\ref{eq:Exc-invariance-To-Spin-Rotation})
in particular. We use whichever reading is convenient, and drop the
tilde on the field from here on.

\emph{The LSDA parent.} The simplest case is collinear LSDA as the
parent. Then Eq.~(\ref{eq:Bxc-general}) becomes 
\begin{equation}
\mathbf{B}_{\mathrm{xc}}=\left(\varepsilon^{\mathrm{LSDA}}_{\mathrm{xc},m}(n,m_{\perp})\,\hat{\mathbf{m}}_{\perp}\right)_{\perp},\qquad\hat{\mathbf{m}}_{\perp}=\frac{\mathbf{m}_{\perp}}{m_{\perp}},\label{eq:SoF-Bxc}
\end{equation}
with $m_{\perp}=|\mathbf{m}_{\perp}|$. \footnote{The unit vector $\hat{\mathbf{m}}_{\perp}$ is undefined where $m_{\perp}$
vanishes, but the product that enters Eq.~(\ref{eq:SoF-Bxc}) is
well behaved there because $\varepsilon^{\mathrm{LSDA}}_{\mathrm{xc},m}$
vanishes linearly or faster as $\mathbf{m}_{\perp}\to0$. } Two features make $\mathbf{B}_{\mathrm{xc}}$ non-parallel to $\mathbf{m}$.
The parent field entering Eq.~(\ref{eq:SoF-Bxc}) is evaluated at
$\mathbf{m}_{\perp}$ and points along $\hat{\mathbf{m}}_{\perp}$
rather than along $\hat{\mathbf{m}}$, and the outer projection is
nonlocal, so it tilts the result further. The local torque density
$\bm{\tau}_{\mathrm{xc}}=\mathbf{m}\times\mathbf{B}_{\mathrm{xc}}$
therefore does not vanish---a key distinction from $\mathbf{B}^{\mathrm{LSDA}}_{\mathrm{xc}}$
of Eq.~(\ref{eq:LSDA-field}). 

\emph{Consequences of the bare-parent approximation.} For the exact
functional of SDFT, and for $\perp$SDFT (Sec.~\ref{sec:transverse-theory}),
the angle $\theta$ between $\mathbf{M}$ and a uniform $\mathbf{B}_{\mathrm{ext}}$
is $180^{\circ}$ and the xc torque vanishes at self-consistency.
The construction of Eq.~(\ref{eq:SoF-Exc-definition}) breaks global
spin-rotation invariance and has no exact energy behind it, so $\bm{\mathcal{T}}_{\mathrm{xc}}$
need not vanish. Since Eq.~(\ref{eq:zero-total-torque-condition})
still holds at stationarity, 
\begin{equation}
\bm{\mathcal{T}}_{\mathrm{xc}}=-\bm{\mathcal{T}}_{\mathrm{ext}}=-\mathbf{M}\times\mathbf{B}_{\mathrm{ext}}.\label{eq:Tau_xc=00003D-Tau_ext_at_stationarity}
\end{equation}
Hence the total moment is no longer collinear with $\mathbf{B}_{\mathrm{ext}}=B\hat{\mathbf{u}}$,
and settles at the angle where $|\mathbf{M}|B\sin\theta=|\bm{\mathcal{T}}_{\mathrm{xc}}|$,
with $\theta\neq0^{\circ},180^{\circ}$. The tilt is measurable. For
a self-consistent solution of a variational energy the only explicit
dependence on $B$ is the Zeeman term, so the Hellmann--Feynman theorem
gives 
\begin{equation}
\left(\frac{\partial E}{\partial B}\right)_{B_{\mathrm{ext}}\ne0}=\int\mathbf{m}\cdot\hat{\mathbf{u}}\,d^{3}r=\mathbf{M}\cdot\hat{\mathbf{u}}=|\mathbf{M}|\cos\theta,\label{eq:Hellmann-Feynman-Theorem}
\end{equation}
so any departure of $\cos\theta$ from $-1$ measures the error of
the bare-parent approximation to $\perp$SDFT. We use Equation~(\ref{eq:Hellmann-Feynman-Theorem})
only with $B_{\mathrm{ext}}\ne0$ to avoid the problems of using the
Hellmann--Feynman theorem at a degeneracy. We use this result below
when we analyze \ce{Mn2} as a test case.

\emph{$N$-representability. }Evaluating the parent at $\mathbf{m}_{\perp}$
violates the $N$-representability bound $|\mathbf{m}|\le n$ \citep{Gontier2013},
since $\mathbf{m}_{\perp}$ decays algebraically while $n$ decays
exponentially. We handle this by saturating the magnetization argument
of the parent at $n$, which leaves Eq.~(\ref{eq:Bxc-general}) and
$\nabla\cdot\mathbf{B}_{\mathrm{xc}}=0$ intact; Appendix~\ref{sec:nrep}
gives the details.

\begin{figure}
\includegraphics[width=1\columnwidth]{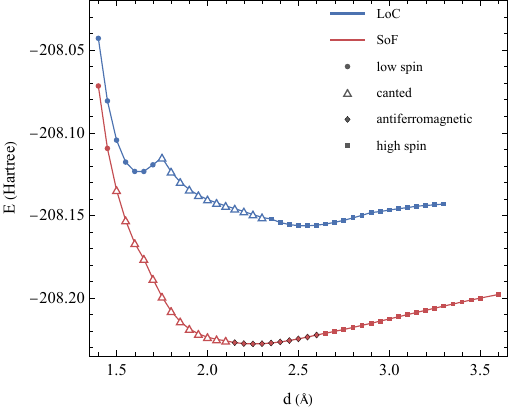}

\caption{\ce{Mn2} potential energy curves for the locally-collinear (LoC)
and source-free (SoF) constructions built on the same parent LSDA
functional. Marker shape indicates the character of the self-consistent
solution: low-spin ($|\mathbf{M}|\approx2$), canted intermediate-moment,
antiferromagnetic ($|\mathbf{M}|\approx0$, the two Mn moments opposed),
or high-spin (aligned, $|\mathbf{M}|>7$). The antiferromagnetic solution
appears only for SoF.}
\label{fig:mn2-pec}
\end{figure}

\section{The manganese dimer as a case study for the bare-parent approximation }\label{sec:manganese-dimer}

Secs.~\ref{sec:impossibility}--\ref{sec:The-SoF-construction}
leave a practical question: what does the bare-parent approximation
do? We check this on the manganese dimer, a prototypical antiferromagnet:
two nearly saturated spin-$5/2$ atoms coupled into a singlet.

\emph{Computational details.} We performed all calculations with a
modified PySCF \citep{Sun2018,Sun2020} code in its generalized Kohn--Sham
(GKS) non-collinear mode. Manganese is described by the Stuttgart
RSC small-core relativistic effective core potential and its accompanying
basis set, leaving the $3s^{2}3p^{6}3d^{5}4s^{2}$ valence shell of
each atom explicit ($86$ contracted Gaussians for the dimer), and
the parent collinear functional is the local spin-density approximation
(Dirac exchange with VWN correlation). The exchange-correlation integrals
use a pruned Treutler--Ahlrichs grid of $2.9\times10^{5}$ points
with Becke partitioning, and self-consistency is converged to $10^{-8}\,\mathrm{E_{h}}$
in the energy and $10^{-4}$ in the orbital gradient.

\emph{Potential energy curves (PECs).} Figure~\ref{fig:mn2-pec}
compares the PECs of the LoC and SoF constructions. We classify each
self-consistent solution by the angle $\alpha$ between the two Mn
atomic moments (obtained from a Mulliken analysis of the converged
density) together with the total-moment magnitude $|\mathbf{M}|$.
Atomic moments and $\alpha$ are properties of the full Kohn--Sham
magnetization $\mathbf{m}[\Phi]$, that is, of the SDFT reading of
Eq.~(\ref{eq:SoF-Exc-definition}) (Sec.~\ref{sec:The-SoF-construction}).
$\alpha=0^{\circ}$ ($180^{\circ}$) means the atomic moments are
parallel (antiparallel); intermediate values describe a canted, non-collinear
arrangement. These parameters help us classify each point with one
of four markers: ($\bullet$) low-spin --- small aligned moment ($|\mathbf{M}|<3$,
$\alpha<45^{\circ}$); ($\blacklozenge$) antiferromagnetic --- near-zero
total moment with opposing atomic moments ($|\mathbf{M}|<0.5$, $\alpha>150^{\circ}$);
($\blacksquare$) high-spin --- large aligned moment ($|\mathbf{M}|>7$,
$\alpha<45^{\circ}$); ($\triangle$) canted --- everything in between.

At the shortest distances ($d\lesssim1.45\,\text{\AA}$) both constructions
yield the same low-spin solution ($\alpha\approx0^{\circ}$, $|\mathbf{M}|\approx2$),
and both eventually enter a canted regime with $\alpha$ well above
$90^{\circ}$. They leave the low-spin solution at very different
distances, however, and beyond it they diverge in how they reach a
large-moment state: SoF passes through an antiferromagnetic regime
while LoC increases spin.

LoC holds the low-spin solution much longer and then jumps in a single
step: at $d=1.70\,\text{\AA}$ we still find it low-spin ($\alpha=0.4^{\circ}$),
while at $d=1.75\,\text{\AA}$ the solution has already switched to
canted ($\alpha=115^{\circ}$, $|\mathbf{M}|$ rising from $2$ to
$3.8$). Because the total energy remains continuous, this abrupt
change of character appears in Fig.~\ref{fig:mn2-pec} as a narrow
$\approx0.23\,\mathrm{eV}$ bump near $d\approx1.7$--$1.75\,\text{\AA}$.
From there $\alpha$ falls smoothly from $115^{\circ}$ to $77.6^{\circ}$
by $d=2.30\,\text{\AA}$ while $|\mathbf{M}|$ grows from $3.8$ to
$7.6$. A second abrupt jump between $d=2.30$ and $2.35\,\text{\AA}$
brings the solution to saturation ($|\mathbf{M}|=10$, $\alpha\approx0^{\circ}$)
-- the high-spin $^{11}\Pi_{u}$ ferromagnet familiar from collinear
LSDA \citep{Ivanov2021}. Beyond this distance LoC remains saturated
and never predicts an antiferromagnetic configuration; its Mulliken
angle stays below $150^{\circ}$ throughout.

SoF makes the same initial shift from low-spin to canted, but much
earlier and far more gradually. Its angle already exceeds $90^{\circ}$
($\alpha=96.6^{\circ}$) at $d=1.5\,\text{\AA}$, so the transition
spreads over a wider interval and produces only a shallow feature
near $d\approx1.65$--$1.7\,\text{\AA}$. Neither $\alpha$ nor $|\mathbf{M}|$
then varies monotonically: $\alpha$ peaks at $139^{\circ}$ by $d=1.65\,\text{\AA}$,
decreases to $126.6^{\circ}$ near $1.90\,\text{\AA}$ and climbs
again, while $|\mathbf{M}|$ rises from about $2$ to a maximum of
$3.2$ at $d=2.00\,\text{\AA}$ before falling back. By $d\approx2.10\,\text{\AA}$
we have $\alpha=136^{\circ}$ and $|\mathbf{M}|=2.9$, and at that
point SoF leaves the LoC path: $|\mathbf{M}|$ collapses to essentially
zero and $\alpha$ locks at $180^{\circ}$. This state, spanning the
range $d\approx2.15$--$2.60\,\text{\AA}$ is an antiferromagnetically
coupled pair of large, oppositely aligned atomic moments that cancel
almost exactly. It is also the region that contains the global minimum
of the SoF curve, at $d_{\mathrm{eq}}\approx2.26\,\text{\AA}$. Finally,
between $d=2.60$ and $2.65\,\text{\AA}$ a third transition carries
the SoF solution to a large aligned moment ($\alpha\approx0^{\circ}$,
$|\mathbf{M}|$ jumping to $8.4$). Unlike the corresponding LoC jump,
this one does not saturate at once. Instead $|\mathbf{M}|$ climbs
continuously from $8.4$ at $d=2.65\,\text{\AA}$ to $9.6$ at $d=3.6\,\text{\AA}$
(the largest distance we scan), never quite reaching the atomic limit
of $10$.

\emph{Bonding.} Table~\ref{tab:mn2-bonding} summarizes the equilibrium
bond length, harmonic frequency, and binding energy for both functionals.
SoF gives the shorter, more strongly bound bond, $d_{\mathrm{eq}}\approx2.26\,\text{\AA}$
against LoC's $2.54\,\text{\AA}$, with a binding energy of $1.3\,\mathrm{eV}$
against $0.8\,\mathrm{eV}$ -- yet it is the softer bond, $\tilde{\nu}\approx180\,\mathrm{cm^{-1}}$
against $210\,\mathrm{cm^{-1}}$ for LoC.

The two constructions give bonds of different orbital character. With
the molecular axis along $x$, the $3d$ shell splits into $\sigma$
($2x^{2}-y^{2}-z^{2}$), $\pi$ ($d_{xy},d_{xz}$) and $\delta$ ($d_{yz},y^{2}-z^{2}$).
The $\sigma$ orbital and one $\delta$ component are both combinations
of $d_{z^{2}}$ and $d_{x^{2}-y^{2}}$, so we separate them using
the axial symmetry, which requires the two $\delta$ populations to
be equal. Evaluating each functional at its own equilibrium gives
Mulliken $3d$ populations $(\sigma,\pi,\pi,\delta,\delta)=(1.38,1.00,1.00,1.02,1.02)$
for SoF at $d=2.25\,\text{\AA}$ and $(1.04,0.99,1.46,1.00,1.00)$
for LoC at $d=2.54\,\text{\AA}$. Both are close to the half-filled
atomic $3d^{5}$ configuration, as one expects for a weakly bound
pair, but they differ in the bonding charge distribution: SoF places
a third of an electron more in $\sigma$, while in LoC the excess
sits in one component of the $\pi$ pair. The SoF bond is thus predominantly
$\sigma$ in character. The strongly unequal $\pi$ populations of
LoC, which would be equal in an axially symmetric $\Sigma$ state,
identify its state as $^{11}\Pi_{u}$: the majority $3d$ shell is
closed, while the single minority electron occupies one component
of the $\pi_{u}$ pair. The imbalance is therefore the signature of
$\Lambda=1$, not of a numerically broken axial symmetry; which of
the two Cartesian components is enriched is arbitrary, the two being
degenerate.

Both lie far from experiment\citep{Baumann1983,Kant1968,Kirkwood1991,Bier1988}
(Table~\ref{tab:mn2-bonding}). Our bonds are a quarter to a third
too short, our frequencies two and a half to three times too stiff,
our wells several times too deep, and both constructions inherit this
from the parent functional\citep{Barborini2016,Yamanaka2007}. At
the experimental bond length of $3.4\,\text{\AA}$ the dimer is held
together largely by long-range correlation between two nearly separate
atomic densities. LSDA cannot describe correlation between nonuniform
densities that barely overlap. What binding it does supply comes from
the region where they do overlap, and there it is too strong, so the
bond closes at too short a range. The trouble is less the correlation
of LSDA than its exchange, since explicit self-interaction corrections
recover near-quantitative agreement with experiment~\citep{Ivanov2021}.

\begin{table}
\begin{tabular}{cccc}
\hline 
 & SoF & LoC & Exp.\tabularnewline
\hline 
character & AFM (BS) & $^{11}\Pi_{u}$ & $^{1}\Sigma^{+}_{g}$\tabularnewline
$r_{e}/\text{\AA}$ & 2.3 & 2.5 & $3.4$\tabularnewline
$\omega_{e}/\mathrm{cm}^{-1}$ & 180 & 210 & $72\pm4$\tabularnewline
$D_{e}/\mathrm{eV}$ & 1.3 & 0.8 & $0.13\pm0.1$\tabularnewline
\hline 
\end{tabular}

\caption{Bond characteristics of \ce{Mn2} from parabolic fits about the minima
of the potential energy curves. Binding energies were obtained from
the energy at each dimer minimum relative to two isolated \ce{Mn}
atoms in the $S=5/2$ ground state. AFM (BS) denotes the broken-symmetry
antiferromagnetic Kohn--Sham solution, which is not a spin eigenstate
and is therefore not assigned a term symbol; it is consistent with
the experimental $^{1}\Sigma^{+}_{g}$ ground state.}
\label{tab:mn2-bonding}
\end{table}

\emph{Exchange coupling.} In order to study the construction's prediction
of magnetic exchange coupling we constrained the total magnetization
$\lvert\mathbf{M}\rvert$ to successive target values $M_{0}$ within
the source-free (SoF) construction (Appendix~\ref{sec:heisenberg})
at the three representative distances $d=2.25$, $2.55$ and $3.00\,\text{\AA}$.
The low-magnetization solutions form a continuous antiferromagnetic
branch, on which we fitted the energies to the Heisenberg form of
Appendix~\ref{sec:heisenberg} 
\begin{equation}
\mathcal{E}\left(M_{0}\right)=\mathrm{const}-Jy,\qquad y=\tfrac{M_{0}}{2}\left(\tfrac{M_{0}}{2}+1\right),\label{eq:constrained-heisenberg-model-Energy}
\end{equation}
by regressing $\mathcal{E}$ linearly against $y$ on the grid $M_{0}=0,1,\dots,5$
in unit steps at each distance. Here $y$ is the lower bound Eq.~(\ref{eq:S2-bound})
on $\langle\hat{\mathbf{S}}^{2}\rangle$, which the model attains
only at even $M_{0}$; the exact constrained minimum of the model
is the piecewise-linear $y_{H}(M_{0})$ of Eq.~(\ref{eq:chord-minimum}),
and we fitted with it as well. We excluded larger values of $M_{0}$,
because there the SCF either fails to converge or jumps discontinuously
onto a ferromagnetically aligned branch (the jump occurs near $M_{0}=6$
at the two shorter distances, and near $M_{0}=9$ at $d=3.00\,\text{\AA}$,
where we could not converge the intervening targets). At $d=3.00\,\text{\AA}$
the antiferromagnetic state is itself only a local minimum: the unconstrained
ground state there is the aligned $|\mathbf{M}|=9.2$ solution of
Fig.~\ref{fig:mn2-pec}, so the fit at that distance characterizes
a metastable branch. As discussed in Appendix~\ref{subsec:Constraint-magnetiz-SDFT},
jumps to lower energy at higher $M_{0}$ are forbidden in the many
body system or in exact SDFT. Their occurrence signifies a deficiency
of the functional, which can be traced back to the parent LSDA, since
the non monotonic behavior also appears in LoC \footnote{In LoC, at $d=2.25\text{\AA}$, $\mathcal{E}\left(M_{0}\right)-\mathcal{E}\left(0\right)=0,-0.33,-0.31,-0.21\,\mathrm{eV}$
for $M_{0}=0,7,8,10$.}.

We did not repeat this constrained procedure for LoC since, as mentioned
above, it does not show an antiferromagnetic branch, even for low
values of $M_{0}$ at any distance we scanned. As a conventional collinear
reference we instead evaluated the Yamaguchi estimator \citep{Yamaguchi1988}
\[
J_{\mathrm{UKS}}=-\frac{E_{\mathrm{HS}}-E_{\mathrm{BS}}}{\left\langle \hat{\mathbf{S}}^{2}\right\rangle _{\mathrm{HS}}-\left\langle \hat{\mathbf{S}}^{2}\right\rangle _{\mathrm{BS}}}
\]
on an unrestricted Kohn--Sham (UKS) high-spin/broken-symmetry pair.
This is an ordinary collinear calculation, not the LoC construction
itself.

Table~\ref{tab:J} gives $J_{\mathrm{UKS}}>0$, a ferromagnetic ground
state. SoF has $J<0$ at every fitted distance, an antiferromagnetic
ground state. Rare-gas matrix measurements find \ce{Mn2} antiferromagnetic,
with $J=-5.8\,\mathrm{cm^{-1}}=-0.72\,\mathrm{meV}$ at $r_{e}$~\citep{Baumann1983,Cheeseman1990,Kirkwood1991}.
UKS is therefore wrong. SoF has the right sign, but its couplings
are too large at the geometries we can converge because $d$ is too
short. $\lvert J_{\mathrm{SoF}}\rvert$ falls by a factor of $4.3$,
from $-12.9$ to $-3.0\,\mathrm{meV}$, between $d=2.55$ and $3.00\,\text{\AA}$.
Continuing at that rate to $r_{e}=3.4\,\text{\AA}$ takes the $-3.0\,\mathrm{meV}$
of Table~\ref{tab:J} to about $-0.8\,\mathrm{meV}$, near the measured
$-0.72\,\mathrm{meV}$.

On the fitted grid, the Heisenberg form Eq.~(\ref{eq:constrained-heisenberg-model-Energy})
reproduces the SoF energies to $\pm2.5\,\mathrm{meV}$ at every point.
Fitting instead with the exact constrained minimum $y_{H}(M_{0})$
changes the couplings by little more than one percent, to $-14.4$,
$-12.7$ and $-3.0\,\mathrm{meV}$ at the three distances, but fits
worse at the two shorter ones, $\pm4.6\,\mathrm{meV}$, while at $3.00\,\text{\AA}$
the two forms are indistinguishable. The constrained Kohn--Sham determinant
varies continuously with $M_{0}$ and shows none of the kinks at even
$M_{0}$ of the multiplet ensemble, so at odd $M_{0}$ its energy
tends to lie below the exact constrained energy of the model, an effect
of the approximate functional. The fit stops for $M_{0}>5$ because
the constrained solution then leaves the antiferromagnetic branch.
A large imposed moment opens a collinear high-spin state, and the
calculation violates the exact monotonicity in $M$ discussed at the
end of Appendix~\ref{subsec:Constraint-magnetiz-SDFT}.\footnote{At $d=2.25$ and $2.55\text{\AA}$ the high-spin solution is only
a local minimum, $82$ and $94\,\mathrm{meV}$ above the AF ground
state. At $d=3\,\text{\AA}$ the $M_{0}=9$ point lies $\sim40\,\mathrm{meV}$
below the unconstrained AF state, consistent with the aligned state
being the ground state at that distance. }.

\begin{table}
\caption{Exchange couplings (meV) and moment ratio $\bar{s}/s$ as a function
of distance $d\,(\text{\protect\AA})$. $R^{2}$ is that of the linear
regression against $y=\tfrac{M_{0}}{2}(\tfrac{M_{0}}{2}+1)$ giving
$J_{\mathrm{SoF}}$ (the regression against $y_{H}$ of Appendix\ref{sec:heisenberg}
gives $-14.4$, $-12.7$ and $-3.0$); $\bar{s}/s$ is estimated from
the unconstrained antiferromagnetic solution (at $d=3.00\,\text{\protect\AA}$
a local minimum); $J_{\mathrm{UKS}}$ comes from a separate collinear
(UKS) calculation.}
\label{tab:J} %
\begin{tabular}{ccccc}
\hline 
$d$ & $J_{\mathrm{SoF}}$ & $R^{2}_{SoF}$ & $\bar{s}/s$ & $J_{\mathrm{UKS}}$\tabularnewline
\hline 
$2.25$ & $-14.6$ & $0.999$ & $1.036$ & $+8.0$\tabularnewline
$2.55$ & $-12.9$ & $0.998$ & $1.019$ & $+12.7$\tabularnewline
$3.00$ & $-3.0$ & $0.992$ & $1.005$ & $+3.5$\tabularnewline
\hline 
\end{tabular}
\end{table}

\emph{Local moments.} A second check uses the Kohn--Sham determinant.
From each unconstrained antiferromagnetic solution we take $c=\langle\hat{\mathbf{S}}^{2}\rangle-M^{2}/4$
and invert Eq.~(\ref{eq:S2KS}) for the effective local moment $\bar{s}=|\langle\hat{\mathbf{S}}_{a}\rangle|$,
\[
\bar{s}^{2}=s(s+1)-c/2\qquad(s=5/2).
\]
The three unconstrained antiferromagnetic states give $\bar{s}=2.590$,
$2.547$ and $2.513$. Each Mn atom is a spin $s=5/2$ site, so the
local moment on an atom should not exceed $2.5$; the Kohn--Sham
values lie a few percent above that bound. The overshoot is smaller
at larger $d$. It also shrinks when the total moment is constrained:
along each $M_{0}$ scan $\bar{s}/s$ falls from $1.036$ to $1.008$
at $2.25\,\text{\AA}$, from $1.019$ to $1.006$ at $2.55\,\text{\AA}$,
and from $1.005$ to $1.001$ at $3.00\,\text{\AA}$.

Putting the energy fit and the local-moment analysis together, we
find that over a limited magnetization window, the SoF Kohn--Sham
solutions for \ce{Mn2} follow the Heisenberg energy law, interpolating
smoothly between the integer-spin points, and closely resemble a pair
of nearly saturated spin-$5/2$ moments.

\begin{figure}
\includegraphics[width=1\columnwidth]{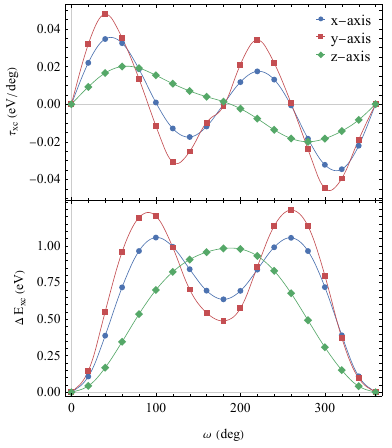}

\caption{The torque \textbf{(top panel)} and the change in the exchange-correlation
energy \textbf{(bottom panel)} of \ce{Mn2} ($d=2.25\,\text{\protect\AA}$)
under a rigid spin rotation of the converged SoF solution by angle
$\omega$ about three orthogonal axes (molecular axis $=\hat{\mathbf{x}}$).
Differentiating a trigonometric fit to the bottom panel reproduces
the top panel to about one percent, as Eq.~(\ref{eq:torque-identity})
requires.}\label{fig:Global-Symmetry-Break}
\end{figure}

\emph{Broken global spin symmetry. }The bare-parent approximation
breaks global spin-rotation invariance and, unlike $\perp$SDFT, has
no exact energy to hide the breaking (Sec.~\ref{sec:transverse-theory}).
We measure its consequences in two ways: by applying a global spin
rotation to a converged solution under zero external field, and by
watching how the converged moment responds to an external field.

From the converged KS solution at $d=2.25\,\text{\AA}$ we rotate
all KS spinors through an angle $\omega$ about each of three orthogonal
axes (by the unitary transformation of Appendix \ref{sec:Orientation-relaxed-self-consist}),
without re-solving the Kohn--Sham equations. The starting solution
was first relaxed with respect to its overall spin orientation by
the procedure of Appendix~\ref{sec:Orientation-relaxed-self-consist},
so that the total torque vanishes at $\omega=0$. This leaves $n(\mathbf{r})$
unchanged and rotates $\mathbf{m}(\mathbf{r})$ rigidly, and we check
if $E_{\mathrm{xc}}(\omega)$ remains constant as required by Eq.~(\ref{eq:Exc-invariance-To-Spin-Rotation}).
As expected, LoC is unchanged to $\mu\mathrm{eV}$ accuracy on every
axis. SoF, however, varies by up to $\sim1\,\mathrm{eV}$ (lower panel
of Fig.~\ref{fig:Global-Symmetry-Break}) --- a small fraction of
the molecular xc energy, $\approx-620\,\mathrm{eV}$, but an energy
variation on the order of a chemical bond. Because the torque vanishes
at the SCF solution, as required by Eq.~(\ref{eq:Tau_xc=00003D-Tau_ext_at_stationarity}),
the SoF curves are stationary at $\omega=0$. Near the SCF point the
violation of Eq.~(\ref{eq:Exc-invariance-To-Spin-Rotation}) is therefore
invisible. The scans about $\hat{\mathbf{y}}$ and about $\hat{\mathbf{z}}$
in the lower panel are visibly different, although the density is
axially symmetric about the internuclear axis $\hat{\mathbf{x}}$.
This is the same symmetry breaking seen from a different angle. In
the converged state the two Mulliken atomic moments ($3.97\,\mu_{B}$
each, $179.8^{\circ}$ apart) point along $\hat{\mathbf{d}}=(0.21,-0.35,0.91)$,
i.e. $25^{\circ}$ from $\hat{\mathbf{z}}$ and nearly perpendicular
to $\hat{\mathbf{x}}$ and $\hat{\mathbf{y}}$. Since the SoF $E_{\mathrm{xc}}$
of Eq.~(\ref{eq:SoF-Exc-definition}) depends on the orientation
of $\mathbf{m}$ relative to the molecular frame, rotating about $\hat{\mathbf{y}}$
and about $\hat{\mathbf{z}}$ sweeps $\hat{\mathbf{d}}$ through different
sets of orientations, and the two scans have no reason to coincide.
A rotation about $\hat{\mathbf{z}}$, the axis closest to $\hat{\mathbf{d}}$,
moves the moments least and gives the smallest torque, whereas rotations
about $\hat{\mathbf{x}}$ and $\hat{\mathbf{y}}$ sweep them through
the full range of orientations; the three curves differ because $\hat{\mathbf{d}}$
is not placed symmetrically with respect to any pair of axes.

Our second test applies a uniform field $\mathbf{B}_{\mathrm{ext}}=B\hat{\mathbf{u}}$
with $B>0$, taking $\hat{\mathbf{u}}$ along the molecular axis $\hat{\mathbf{x}}$
and along $\hat{\mathbf{y}}$ and $\hat{\mathbf{z}}$, and follows
the tilt $\theta$ between $\mathbf{M}$ and $\mathbf{B}_{\mathrm{ext}}$.
We extract $\theta$ from 
\[
\cos\theta=\frac{\mathbf{M}\cdot\hat{\mathbf{u}}}{\left|\mathbf{M}\right|}=\frac{1}{\left|\mathbf{M}\right|}\frac{\partial E}{\partial B},
\]
which combines the two routes of Sec.~\ref{sec:The-SoF-construction}
--- the converged moment and the energy slope of Eq.~(\ref{eq:Hellmann-Feynman-Theorem})
--- and we require the two to agree, as the Hellmann--Feynman theorem
requires.\footnote{Only states with $M(B=0)\ne0$ can serve: in the antiferromagnetic
states $\mathbf{M}$ vanishes at zero field, $\theta$ is undefined
there, and both $\mathbf{M}$ and $\bm{\mathcal{T}}_{\mathrm{xc}}$
go to zero with $B$.} At very weak fields the Hellmann--Feynman condition was not satisfied
with sufficient accuracy because of incomplete SCF convergence; we
therefore did not use fields weaker than $1.5\times10^{-5}\,E_{h}/\mu_{B}$.

\begin{figure}
\includegraphics[width=1\columnwidth]{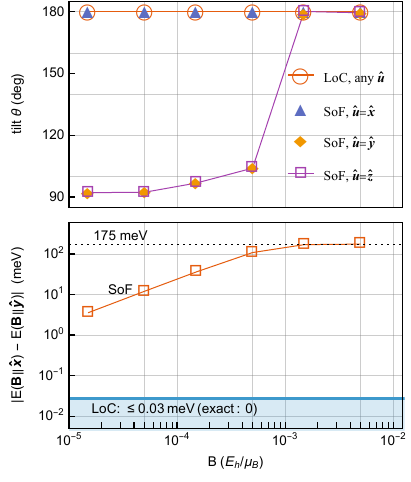}

\caption{Response of $\mathrm{Mn}_{2}$ ($d=3.00\,\text{\protect\AA}$) to
a uniform field $\mathbf{B}=B\hat{\mathbf{u}}$.\textbf{ (Upper panel)}
Tilt $\theta$ between the total moment $\mathbf{M}$ and $\mathbf{B}$,
applied along the molecular axis ($\hat{\mathbf{u}}=\hat{\mathbf{x}}$)
or perpendicular to it ($\hat{\mathbf{u}}=\hat{\mathbf{y}},\hat{\mathbf{z}}$).\textbf{
(Lower panel)} Energy difference $\Delta E(B)=E(\hat{\mathbf{y}},B)-E(\hat{\mathbf{x}},B)$.
The shaded band marks the LoC result, consistent with zero.}\label{fig:mn2-field}
\end{figure}

Figure~\ref{fig:mn2-field} (upper panel) shows, at $d=3.00\,\text{\AA}$,
the tilt $\theta$ as a function of the external field $B$ as it
changes over two and a half decades, $B=1.5\times10^{-5}$ to $5\times10^{-3}\,E_{h}/\mu_{B}$.
As above, we use only states where the zero field magnetization $M\left(0\right)$
is non-zero ($9.6\mu_{B}$ for SoF and $10\mu_{B}$ for LoC). The
LoC tilt is $\theta=180^{\circ}$ for all three field directions and
every $B$: $\mathbf{M}$ stays antiparallel to $\mathbf{B}_{\mathrm{ext}}$,
as a rotationally invariant functional requires. SoF behaves similarly
only when the field is aligned with the molecular axis ($\hat{\mathbf{x}}$).
A perpendicular field gives $\theta\simeq92^{\circ}$ for weak fields.
This means that the moment lies almost entirely along the molecule
instead of along the field. The tilt drifts only slowly, from $92^{\circ}$
to $105^{\circ}$, as $B$ rises to $5\times10^{-4}\,E_{h}/\mu_{B}$,
and then jumps to $180^{\circ}$ by $B=1.5\times10^{-3}\,E_{h}/\mu_{B}$,
where it stops changing. Equation~(\ref{eq:Tau_xc=00003D-Tau_ext_at_stationarity})
accounts for the shape: a nonvanishing exchange-correlation torque
$\left|\bm{\mathcal{T}}_{\mathrm{xc}}\right|=MB\sin\theta$ holds
the moment to the molecular axis --- against the pull of the external
field.

Let $E(\hat{\mathbf{u}},B)$ denote the converged total energy in
the field $\mathbf{B}_{\mathrm{ext}}=B\hat{\mathbf{u}}$, and $\mathbf{M}(\hat{\mathbf{u}},B)$
the corresponding total moment. The lower panel of Fig.~\ref{fig:mn2-field}
shows the energy difference between the two field directions at the
same $B$, 
\[
\Delta E(B)=E(\hat{\mathbf{y}},B)-E(\hat{\mathbf{x}},B).
\]
For a rotationally invariant functional, including the exact one,
the energy in a uniform field depends only on $B$, so $\Delta E(B)=0$
at every $B$. The same holds for $\perp$SDFT, whose functionals
are not invariant but whose energy is exact (Sec.~\ref{sec:transverse-theory}).
A nonzero $\Delta E$ is therefore an error, and to see what kind
of error we split it into two parts. Write the converged energy as
$E(\hat{\mathbf{u}},B)=E_{\mathrm{int}}(\hat{\mathbf{u}},B)+\mathbf{M}\cdot\mathbf{B}_{\mathrm{ext}}$,
where $E_{\mathrm{int}}$ is the Kohn--Sham energy without the Zeeman
term. Then 
\begin{widetext}
\[
\Delta E(B)=\underbrace{E_{\mathrm{int}}(\hat{\mathbf{y}},B)-E_{\mathrm{int}}(\hat{\mathbf{x}},B)}_{\text{magnetic anisotropy}}+\underbrace{B\left[\mathbf{M}(\hat{\mathbf{y}},B)\cdot\hat{\mathbf{y}}-\mathbf{M}(\hat{\mathbf{x}},B)\cdot\hat{\mathbf{x}}\right]}_{\text{Zeeman difference}}.
\]
\end{widetext}

The first term is the energy cost of the moment's orientation relative
to the molecule, a magnetic anisotropy; the second is the difference
in Zeeman energy, nonzero only if the moment aligns differently with
the field in the two cases. In an invariant theory both differences
vanish separately. For SoF, below saturation, the Zeeman difference
dominates since $\mathbf{M}$ is nearly perpendicular to $\hat{\mathbf{y}}$
while being antiparallel to $\hat{\mathbf{x}}$. Once both directions
saturate, the Zeeman difference cancels, revealing the anisotropy.
For any rotationally invariant functional, including the exact xc
functional, the total energy and the Zeeman term depend only on $B$
and not on $\hat{\mathbf{u}}$. The same holds for $\perp$SDFT, whose
functionals are not invariant but whose energy is exact (Sec.~\ref{sec:transverse-theory}).
Therefore, in this case both the anisotropy and the Zeeman difference
vanish at every field, and with them $\Delta E(B)$. The LoC differences
in the panel confirm this: they lie inside the shaded band at the
bottom, effectively zero to our numerical accuracy. However, for SoF
we notice a spurious $\Delta E$ of already $3.5\,\mathrm{meV}$ at
the weakest field we applied. It grows roughly linearly with $B$
and then levels off near $175\,\mathrm{meV}$ once the tilt reaches
its ``correct'' value $180^{\circ}$. That plateau is therefore
the spurious anisotropy barrier. 

In absolute terms this anisotropy is a small energy, and it is reached
only at fields of order $10^{-3}\,E_{h}/\mu_{B}$ ($\approx500\,\mathrm{T}$),
far beyond current laboratory reach. It is nonetheless an artifact
more than an order of magnitude larger than the computed exchange
couplings of Table~\ref{tab:J}. The $\hat{\mathbf{x}}-\hat{\mathbf{y}}$
anisotropy differs from that of $\hat{\mathbf{x}}-\hat{\mathbf{z}}$
by at most $0.12\,\mathrm{meV}$, small but above our numerical precision.
Both the anisotropy and the tilt are properties of the bare-parent
approximation, not of the source-free condition: the applied field
is uniform and hence source-free, so it lies within the domain of
$\perp$SDFT, which for the same field gives $\theta=180^{\circ}$
and no anisotropy (Sec.~\ref{sec:transverse-theory}).

\section{Conclusions}\label{sec:conclusions}

A source-free exchange-correlation magnetic field is not an exact
property of SDFT: no functional invariant under global spin rotations
has one. It is an exact property of $\perp$SDFT, the transverse theory
of Ref.~\citep{Sharma2018}, but there it belongs to a different
functional, $\tilde{E}_{\mathrm{xc}}$, which contains part of the
non-interacting kinetic energy, and it comes at the price of the longitudinal
magnetization, which the Kohn--Sham system of $\perp$SDFT does not
deliver. Exact $\perp$SDFT reproduces the energy, density and transverse
magnetization of SDFT; the condition itself changes nothing, and whatever
a source-free field changes in practice is a property of the approximation
that imposes it. We examined the simplest one, the bare-parent approximation
of Eq.~(\ref{eq:SoF-Exc-definition}), which evaluates a parent functional
at the transverse magnetization. Its $\mathbf{B}_{\mathrm{xc}}$ is
divergence-free, remains a functional derivative, and exerts the local
torques missing from a locally collinear parent, while reducing to
that parent for the homogeneous electron gas.

On $\mathrm{Mn}_{2}$, with LSDA as the parent, the construction reproduces
the magnetic improvement reported with source-free fields~\citep{Sharma2018,Krishna2019,Dewhurst2018,Hawkhead2026,Moore2025}:
a broken-symmetry antiferromagnet, consistent with the experimental
$^{1}\Sigma^{+}_{g}$ ground state and built from nearly saturated
spin-$5/2$ local moments that follow a Heisenberg energy law. The
exchange coupling has the correct sign and, extrapolated to the experimental
bond length, roughly the correct size. The locally collinear control
finds instead the high-spin $^{11}\Pi_{u}$ ferromagnet, no antiferromagnetic
state at any distance, and the wrong sign of $J$. Bond lengths, vibrational
frequencies, and binding energies remain poor in both constructions.
That failure is due to the limitations of the LSDA parent and not
to the construction.

The bare-parent approximation breaks global spin-rotation invariance
and, unlike $\perp$SDFT, has no exact energy behind it, so the breaking
shows. A rigid spin rotation of the converged solution changes the
energy by up to $\sim1\,\mathrm{eV}$, where the change should be
zero. The functional produces a uniaxial magnetic anisotropy of about
$175\,\mathrm{meV}$, more than an order of magnitude larger than
the computed exchange couplings; the exact anisotropy is zero in the
absence of spin--orbit coupling, and so is that of $\perp$SDFT.
In a weak field the magnetization does not align with $-\mathbf{B}_{\mathrm{ext}}$;
on \ce{Mn2} in a weak perpendicular field it is nearly perpendicular
to $\mathbf{B}_{\mathrm{ext}}$. The same breaking will affect the
electron dynamics: the zero xc-torque theorem is restored only at
the SCF minimum, so real-time propagation away from that point acquires
a spurious contribution to $d\langle\hat{\mathbf{S}}\rangle/dt$,
as in Ref.~\citep{BologaUllrich2024}. Desmarais et al.\citep{Desmarais2026}
have recently shown that within spin-current DFT, local $U(1)\times SU(2)$
invariance yields a local xc-torque identity under which the xc torques
shape the ground state but drop out of the equation of motion of the
magnetization; this addresses the xc-torque problem while respecting
the symmetry that a source-free condition breaks. The magnetic gain
on \ce{Mn2} and the directional errors have the same origin, the
approximation; both are absent from exact $\perp$SDFT, which reproduces
the energy of SDFT. Whether a better parent than LSDA, or an approximation
to $\tilde{E}_{\mathrm{xc}}$ that retains part of the minimization
over the longitudinal magnetization, can keep the antiferromagnet
and the sign of $J$ while reducing the error, is left open.
\begin{acknowledgments}
This work was supported by the Israel Science Foundation, Grant No.~ISF-1153/23.
The authors used Anthropic's Claude to help trace the anomalous field
response of the source-free construction to the breaking of global
spin-rotation symmetry, in implementing the fix into the PySCF code,
and in the editing of the manuscript. The authors verified all results
and take full responsibility for the content.
\end{acknowledgments}

\appendix

\section{The transverse projection}\label{sec:TheTransverseProjection}

The Helmholtz theorem states that any vector field $\mathbf{a}(\mathbf{r})$
can be decomposed into a sum of longitudinal $\mathbf{a}_{\parallel}$
and transverse $\mathbf{a}_{\perp}$ parts, where $\nabla\times\mathbf{a}_{\parallel}=0$
and $\nabla\cdot\mathbf{a}_{\perp}=0$. The theorem also gives an
algorithm, since $\mathbf{a}_{\parallel}$ is a conservative field
it can be determined as the gradient of a potential $\mathbf{a}_{\parallel}=\nabla\chi$
determined from the Poisson equation $\nabla^{2}\chi=\nabla\cdot\mathbf{a}$.
Here, $\nabla\cdot\mathbf{a}$ is the ``source distribution'' of
$\mathbf{a}$ just as the electric charge $\rho$ is the source of
the Maxwell's electric field $\nabla\cdot\mathbf{E=\rho}.$ If this
is a source-free field then $\nabla\cdot\mathbf{a}=0$, $\mathbf{a}_{\parallel}=0$
and $\mathbf{a}$ has no longitudinal component. Once $\mathbf{a}_{\parallel}$
is known the transverse part is immediately determined: $\mathbf{a}_{\perp}=\mathbf{a}-\mathbf{a}_{\parallel}$.

For localized systems the vector field decays sufficiently fast to
zero allowing the use of a Coulomb-like integral to determine the
potential from the sources: $\chi(\mathbf{r})=-\frac{1}{4\pi}\int\frac{\nabla'\cdot\mathbf{a}(\mathbf{r}')}{|\mathbf{r}-\mathbf{r}'|}\,d^{3}r'$.
The longitudinal component is therefore the analog of an electric
field: $\mathbf{a}_{\parallel}(\mathbf{r})=\frac{1}{4\pi}\int\frac{\nabla'\cdot\mathbf{a}(\mathbf{r}')}{|\mathbf{r}-\mathbf{r}'|^{2}}\frac{\mathbf{r}-\mathbf{r}'}{|\mathbf{r}-\mathbf{r}'|}\,d^{3}r'$.
Let us discuss the asymptotic form of this field through the low order
moments of its source charge distribution $\nabla'\cdot\mathbf{a}(\mathbf{r}')$
. The zeroth moment is the charge and it is a volume integral of the
distribution. Because $\mathbf{a}$ is localized this charge is zero
\footnote{From Gauss' theorem: $\int\nabla'\cdot\mathbf{a}(\mathbf{r}')\,d^{3}r'=\ointop_{S}\mathbf{a}\cdot d^{2}\mathbf{s}=0$,
where $S$ is a surface at infinity.} The next moment is the dipole and that, by integration by parts,
evaluates to the negative of the volume integral of $\mathbf{a}$
itself: 
\begin{equation}
\mathbf{p}=\int\nabla'\cdot\mathbf{a}(\mathbf{r})\mathbf{r}\,d^{3}r=-\int\mathbf{a}(\mathbf{r})\,d^{3}r.\label{eq:Field-dipole}
\end{equation}
So the potential decays as $\chi\approx-\frac{\mathbf{p}\cdot\mathbf{r}}{4\pi r^{3}}$
for large $r$ and the longitudinal component of the field is that
of this dipole. Since $\mathbf{a}$ itself decays faster than any
power, in the far field $\mathbf{a}_{\perp}=\mathbf{a}-\mathbf{a}_{\parallel}\approx-\mathbf{a}_{\parallel}$,
so
\begin{align}
\mathbf{a}_{\parallel}(\mathbf{r}) & \approx-\mathbf{a}_{\perp}(\mathbf{r})\approx\frac{3\left(\mathbf{p}\cdot\hat{\mathbf{r}}\right)\hat{\mathbf{r}}-\mathbf{p}}{4\pi r^{3}},\label{eq:dipole-field}
\end{align}
and hence $\left|\mathbf{a}_{\parallel}\right|\approx\left|\mathbf{a}_{\perp}\right|\approx\frac{\left|\mathbf{p}\right|}{4\pi r^{3}}\sqrt{1+3\cos^{2}\vartheta}$,
with $\vartheta$ the angle between $\mathbf{r}$ and $\mathbf{p}$.
The angular factor lies between $1$ and $2$ and never vanishes,
so the tail is present in every direction. Note even when $\mathbf{a}$
is totally localized, both $\mathbf{a}_{\parallel}$ and $\mathbf{a}_{\perp}$
are not: they decay algebraically to zero.

In the Fourier representation, $\tilde{\mathbf{a}}(\mathbf{k})=\int\mathbf{a}(\mathbf{r})e^{-i\mathbf{k}\cdot\mathbf{r}}\,d^{3}r$,
the decomposition is algebraic, 
\begin{equation}
\tilde{\mathbf{a}}_{\parallel}(\mathbf{k})=\hat{\mathbf{k}}\bigl(\hat{\mathbf{k}}\cdot\tilde{\mathbf{a}}(\mathbf{k})\bigr),\qquad\tilde{\mathbf{a}}_{\perp}(\mathbf{k})=\bigl(\mathbb{I}-\hat{\mathbf{k}}\hat{\mathbf{k}}^{\mathsf{T}}\bigr)\tilde{\mathbf{a}}(\mathbf{k}),\label{eq:S-fourier}
\end{equation}
with $\hat{\mathbf{k}}=\mathbf{k}/|\mathbf{k}|$.

For lattice-periodic fields we Fourier-series expand $\mathbf{a}\left(\mathbf{r}\right)=\sum_{\mathbf{G}}\tilde{\mathbf{a}}(\mathbf{G})e^{i\mathbf{G}\cdot\mathbf{r}}$
in the reciprocal-lattice vectors $\mathbf{G}$ and we apply the projector
of Eq.~(\ref{eq:S-fourier}) with $\hat{\mathbf{k}}\to\hat{\mathbf{G}}$.
The uniform component $\tilde{\mathbf{a}}(\mathbf{G}=0)$ has zero
divergence and zero curl and is naturally assigned to the transverse
part of $\mathbf{a}$: 
\[
\tilde{\mathbf{a}}_{\perp}(\mathbf{G}=0)=\tilde{\mathbf{a}}(\mathbf{G}=0),\qquad\tilde{\mathbf{a}}_{\parallel}(\mathbf{G}=0)=0,
\]
consistent with the localized case, where the sources $\nabla\cdot\mathbf{a}$
generate the longitudinal part exclusively and a uniform field has
none.

The Helmholtz map $\mathscr{H}$ taking $\mathbf{a}$ to its transverse
part $\mathbf{a}_{\perp}$ obeys: $\mathscr{H}^{2}=\mathscr{H}$ (once
the field is source-free there is no longitudinal part any more).
From Eq.~(\ref{eq:S-fourier}), in Fourier space $\mathscr{H}$ acts
at each $\mathbf{k}$ (or $\mathbf{G}$) as $\mathbb{I}-\hat{\mathbf{k}}\hat{\mathbf{k}}^{\mathsf{T}}$
showing that it is real and symmetric, i.e. $\mathscr{H}^{\dagger}=\mathscr{H}$.
These two properties mean that $\mathscr{H}$ is a self-adjoint projection
with respect to the inner product $\langle\mathbf{a},\mathbf{b}\rangle=\int\mathbf{a}\cdot\mathbf{b}\,d^{3}r$.
One important conclusion is the following identity: 
\begin{equation}
\langle\mathbf{a},\mathbf{b}_{\perp}\rangle=\langle\mathbf{a},\mathscr{H}\mathbf{b}\rangle=\langle\mathscr{H}\mathbf{a},\mathbf{b}\rangle=\langle\mathbf{a}_{\perp},\mathbf{b}\rangle.\label{eq:symmetricHelmholtzProj}
\end{equation}
Furthermore: 
\begin{equation}
\langle\mathbf{a}_{\parallel},\mathbf{a}_{\perp}\rangle=0.\label{eq:TransLong-Orthogonality}
\end{equation}

\begin{figure}
\includegraphics[width=1\columnwidth]{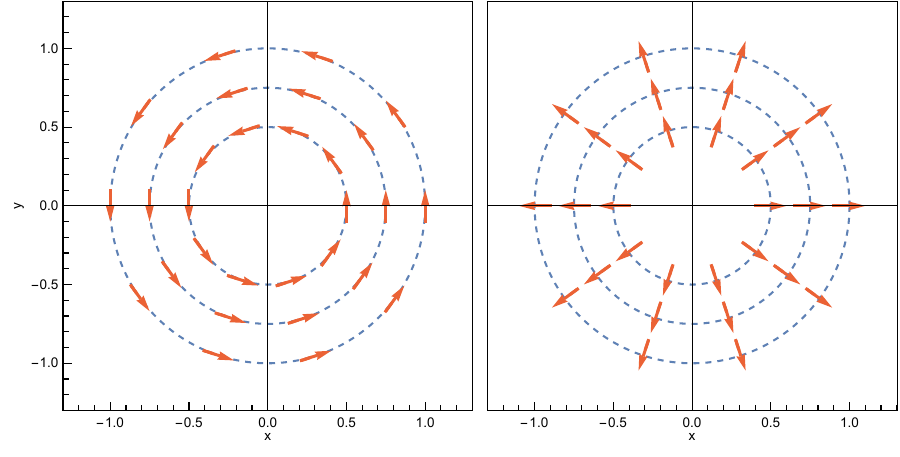}

\caption{A demonstration how a clockwise global spin-rotation by $90^{0}$
converts a source-free field, where all field lines are loops (left)
into a field where all field lines are rays emanating from a source
at the origin (right).}\label{fig:curldiv}
\end{figure}

\section{Global rotations and Helmholtz components}\label{sec:Global-rotations-mix}

The Helmholtz decomposition is not preserved when the vector index
of the field is globally rotated. Let $\mathcal{R}_{\bm{\omega}}$
be a constant rotation by an infinitesimal angle $\omega$ about the
axis $\hat{\mathbf{n}}$, with rotation vector $\bm{\omega}=\omega\hat{\mathbf{n}}$.
When $\mathcal{R}_{\bm{\omega}}$ operates on a vector field it operates
at each point in the same way: 
\begin{equation}
\mathcal{R}_{\bm{\omega}}\mathbf{a}=\mathbf{a}+\bm{\omega}\times\mathbf{a}+O\left(\omega^{2}\right).\label{eq:infinitesimal-rotation}
\end{equation}
Since $\bm{\omega}$ is constant, $\nabla\cdot(\bm{\omega}\times\mathbf{a})=-\bm{\omega}\cdot(\nabla\times\mathbf{a})$
and $\nabla\times(\bm{\omega}\times\mathbf{a})=\bm{\omega}(\nabla\cdot\mathbf{a})-(\bm{\omega}\cdot\nabla)\mathbf{a}$,
whence, up to $O\left(\omega^{2}\right)$,
\begin{align}
\nabla\cdot\left(\mathcal{R}_{\bm{\omega}}\mathbf{a}\right) & =\nabla\cdot\mathbf{a}-\bm{\omega}\cdot\left(\nabla\times\mathbf{a}\right)+O\left(\omega^{2}\right),\label{eq:div-rotation}\\
\nabla\times\left(\mathcal{R}_{\bm{\omega}}\mathbf{a}\right) & =\nabla\times\mathbf{a}+\bm{\omega}\left(\nabla\cdot\mathbf{a}\right)-(\bm{\omega}\cdot\nabla)\mathbf{a}+O\left(\omega^{2}\right)\label{eq:curl-rotation}
\end{align}
The two Helmholtz components behave differently under infinitesimal
rotation. To first order $\bm{\omega}\times\mathbf{a}_{\parallel}$
is transverse, since its divergence is $-\bm{\omega}\cdot(\nabla\times\mathbf{a}_{\parallel})=0$:
rotating a longitudinal field produces only a transverse part. Rotating
a transverse field produces both parts. A finite angle global rotation
mixes both components: the curl into the divergence and the divergence
into the curl. A graphic demonstration of this is given in Fig. \ref{fig:curldiv}
for a right angle spin rotation. From Eq.~(\ref{eq:div-rotation})
it follows that a source-free field can stay source-free under all
rotations $\mathcal{R}_{\bm{\omega}}$ only if $\nabla\times\mathbf{a}=0$.

The curl is also not covariant under global spin rotations. Applying
Eq.~(\ref{eq:infinitesimal-rotation}) to $\nabla\times\mathbf{a}$
and using Eq.~(\ref{eq:curl-rotation}), 
\begin{equation}
\nabla\times\left(\mathcal{R}_{\bm{\omega}}\mathbf{a}\right)-\mathcal{R}_{\bm{\omega}}\left(\nabla\times\mathbf{a}\right)=\bm{\omega}\left(\nabla\cdot\mathbf{a}\right)-\nabla(\bm{\omega}\cdot\mathbf{a})+O\left(\omega^{2}\right),\label{eq:curl-covariance}
\end{equation}
which vanishes for every $\bm{\omega}$ if and only if $\mathbf{a}$
is uniform \footnote{Take $\bm{\omega}=\hat{\mathbf{e}}_{k}$. Then $\hat{\mathbf{e}}_{k}\nabla\cdot\mathbf{a}=\nabla a_{k}$.
For $j\ne k$ $\partial_{j}a_{k}=0$. For $j=k$ $\partial_{k}a_{k}=\nabla\cdot\mathbf{a}$
(no summation). Summing over k gives $\nabla\cdot\mathbf{a}=3\nabla\cdot\mathbf{a}$,
which means $\nabla\cdot\mathbf{a}=0$ and thus $\partial_{k}a_{k}=0$
(no summation). Thus $\partial_{j}a_{k}=0$ everywhere for all $j$
and $k$. $\mathbf{a}$ is either a uniform vector field. }. The right-hand side depends on $\nabla\cdot\mathbf{a}$ and on $\bm{\omega}\cdot\mathbf{a}$,
neither of which is fixed solely by $\nabla\times\mathbf{a}$. 

Two consequences follow. First, the curl is not covariant: rotating
$\mathbf{a}$ and then taking the curl differs from rotating $\nabla\times\mathbf{a}$.
Second, and stronger, the curl of the rotated field is not determined
by the curl of the original one. Take $\mathbf{a}$ and $\mathbf{a}'=\mathbf{a}+\nabla\chi$,
which have the same curl. Applying the equation above to each and
subtracting, the two rotated fields have curls that differ by 
\begin{equation}
\nabla\times(\mathcal{R}_{\bm{\omega}}\mathbf{a}')-\nabla\times(\mathcal{R}_{\bm{\omega}}\mathbf{a})=\bm{\omega}\,\nabla^{2}\chi-\nabla(\bm{\omega}\cdot\nabla\chi)+O(\omega^{2}),
\end{equation}
which is nonzero unless $\nabla\chi$ is uniform. So a functional
$F[\nabla\times\mathbf{a}]$ takes one value on $\mathbf{a}$ and
$\mathbf{a}'$ but, after the same rotation, two different values.
A global spin rotation therefore is not a well-defined operation on
the variable $\nabla\times\mathbf{a}$, and the invariance condition
Eq.~(\ref{eq:Exc-invariance-To-Spin-Rotation}) cannot be written
for a functional of it (Sec.~\ref{sec:transverse-theory}).

The same holds for the transverse part. $\mathbf{a}$ and $\mathbf{a}'=\mathbf{a}+\nabla\chi$
have the same $\mathbf{a}_{\perp}$. After rotation they differ by
$\mathcal{R}_{\bm{\omega}}\nabla\chi=\nabla\chi+\bm{\omega}\times\nabla\chi+O(\omega^{2})$,
whose first-order term is transverse, since $\nabla\cdot(\bm{\omega}\times\nabla\chi)=-\bm{\omega}\cdot(\nabla\times\nabla\chi)=0$.
Hence 
\begin{equation}
(\mathcal{R}_{\bm{\omega}}\mathbf{a}')_{\perp}-(\mathcal{R}_{\bm{\omega}}\mathbf{a})_{\perp}=\bm{\omega}\times\nabla\chi+O(\omega^{2}),
\end{equation}
which is nonzero unless $\nabla\chi$ is parallel to $\bm{\omega}$.
So a functional $F[\mathbf{a}_{\perp}]$ also takes one value on $\mathbf{a}$
and $\mathbf{a}'$ but two different values after the same rotation,
and a global spin rotation is not a well-defined operation on $\mathbf{a}_{\perp}$
either. The longitudinal part fares no better: $\mathbf{a}$ and $\mathbf{a}+\mathbf{t}$
with $\mathbf{t}$ transverse share $\mathbf{a}_{\parallel}$, and
after rotation their longitudinal parts differ by $(\bm{\omega}\times\mathbf{t})_{\parallel}$,
nonzero in general. Only the full field $\mathbf{a}$ carries an action
of the rotation group. 

A functional of the full field built from the curl, $G[\mathbf{a}]=F[\nabla\times\mathbf{a}]$
or $G[\mathbf{a}]=F[\mathbf{a}_{T}]$ does admit the rotation, but
it cannot be invariant: $G$ is blind to $\nabla\chi$, invariance
would make it blind to $\mathcal{R}^{\mathsf{T}}_{\bm{\omega}}\nabla\chi$
for every $\bm{\omega}$ as well, and these together span all fields,
so $G$ would be constant. This is the content of Sec.~\ref{sec:impossibility}.

\section{ Heisenberg model under constrained magnetization}\label{sec:heisenberg}

\subsection{The spin-constrained Heisenberg model}

The Heisenberg model stipulates that the low-energy spin spectrum
of the dimer is captured by a Hilbert space of states describing two
atomic spins, $\hat{\mathbf{S}}_{1}$, $\hat{\mathbf{S}}_{2}$, each
with a sharp value for $\hat{S}^{2}_{a}$, namely $s\left(s+1\right)$
(but $S^{z}_{1}$ and $S^{z}_{2}$ can vary). This space has the dimension
$\left(2s+1\right)^{2}$. The spins can only interact through a spin-rotation
invariant, the simplest of which is $\hat{H}_{H}=-2J\hat{\mathbf{S}}_{1}\cdot\hat{\mathbf{S}}_{2}$,
with $J$ the exchange coupling coefficient. The Hamiltonian eigenstates
are also eigenstates associated with the total spin $\hat{\mathbf{S}}=\hat{\mathbf{S}}_{1}+\hat{\mathbf{S}}_{2}$,
described by the quantum number $S$, having, in our case, the values
$S=0,\dots,2s$ in integer steps, with $\hat{\mathbf{S}}^{2}$ having
the eigenvalues $S\left(S+1\right)$. Furthermore, $\hat{\mathbf{S}}^{2}=2s\left(s+1\right)+2\hat{\mathbf{S}}_{1}\cdot\hat{\mathbf{S}}_{2}$,
so the Hamiltonian is $\hat{H}_{H}=-J\left(\hat{\mathbf{S}}^{2}-2s\left(s+1\right)\right)$
and its eigenvalues are $E_{S}=-J\left[S\left(S+1\right)-2s\left(s+1\right)\right]$,
each with degeneracy of $2S+1$. The energy differences are controlled
by $J$: $\Delta E_{S-1\to S}=-2JS$. Positive (negative) values of
$J$ favor high (low) $S$ states.

For any wavefunction $\Psi$ (whether or not an eigenstate), 
\begin{equation}
E_{H}=\left\langle \Psi\left|\hat{H}_{H}\right|\Psi\right\rangle =-J\left(\left\langle \Psi\left|\hat{\mathbf{S}}^{2}\right|\Psi\right\rangle -2s\left(s+1\right)\right).\label{eq:affine}
\end{equation}

For comparing with SDFT calculations we search for the minimal energy
of the Heisenberg model under the corresponding constraint $\left|\left\langle \hat{\mathbf{S}}\right\rangle \right|=M_{0}/2$.
When $J<0$ we search for the minimal value of $\left\langle \hat{\mathbf{S}}^{2}\right\rangle $.
Using the extended uncertainty principle, it is possible to show \footnote{For this, we write $\left\langle \hat{\mathbf{S}}^{2}\right\rangle =\left|\left\langle \hat{\mathbf{S}}\right\rangle \right|^{2}+\sum_{a}\left[\left\langle \hat{S}^{2}_{a}\right\rangle -\left\langle \hat{S}_{a}\right\rangle ^{2}\right]$,
the sum running over three orthogonal directions. Taking $z$ along
$\left\langle \hat{\mathbf{S}}\right\rangle $, so that $\left\langle \hat{S}_{z}\right\rangle =M_{0}/2$,
and denoting the two transverse components by $\hat{S}_{\perp1}$
and $\hat{S}_{\perp2}$, the uncertainty relation $\mathrm{Var}(\hat{S}_{\perp1})\mathrm{Var}(\hat{S}_{\perp2})\ge\tfrac{1}{4}\left\langle \hat{S}_{z}\right\rangle ^{2}$
together with the inequality between the arithmetic and geometric
means gives $\mathrm{Var}(\hat{S}_{\perp1})+\mathrm{Var}(\hat{S}_{\perp2})\ge2\sqrt{\mathrm{Var}(\hat{S}_{\perp1})\mathrm{Var}(\hat{S}_{\perp2})}\ge M_{0}/2$,
and since $\mathrm{Var}(\hat{S}_{z})\ge0$, $\left\langle \hat{\mathbf{S}}^{2}\right\rangle \ge\frac{M_{0}}{2}\left(\frac{M_{0}}{2}+1\right)$.} that 
\begin{equation}
\left\langle \hat{\mathbf{S}}^{2}\right\rangle \ge\frac{M_{0}}{2}\left(\frac{M_{0}}{2}+1\right),\label{eq:S2-bound}
\end{equation}
with equality for the states $\left|S,S\right\rangle $ at $M_{0}=2S$.
The bound is attained only there. In the model space the total spin
takes the integer values $S=0,\dots,2s$, so a state with $\left\langle \hat{S}_{z}\right\rangle =q\equiv M_{0}/2$
lying between the integers $k$ and $k+1$ must mix multiplets. Expanding
it in sectors of total spin with weights $p_{S}$, we have $\left\langle \hat{\mathbf{S}}^{2}\right\rangle =\sum_{S}p_{S}S(S+1)$
and $\sum_{S}p_{S}S\ge q$; since $S(S+1)$ is convex in $S$, the
constrained minimum is the linear interpolation between neighboring
integers,\begin{widetext}
\begin{equation}
y_{H}\left(M_{0}\right)=\left(1+k-q\right)k\left(k+1\right)+\left(q-k\right)\left(k+1\right)\left(k+2\right),\qquad k=\lfloor q\rfloor,\label{eq:chord-minimum}
\end{equation}
\end{widetext}reached by superposing $\left|k,k\right\rangle $ and
$\left|k+1,k+1\right\rangle $. 

When $J>0$ we want to maximize $\left\langle \hat{\mathbf{S}}^{2}\right\rangle $.
But for any $M_{0}\le4s$ we can saturate it at $\left(2s\right)\left(2s+1\right)$
by superposing members of the $S=2s$ multiplet. Thus, for $J>0$
the energy does not depend on $M_{0}$. These considerations are summarized
as: 
\begin{equation}
\mathcal{E}_{H}\left(M_{0}\right)=\begin{cases}
\mathrm{const}, & J>0,\\[2pt]
\mathrm{const}-J\,y_{H}\left(M_{0}\right), & J<0.
\end{cases}\label{eq:minHeisenbergE}
\end{equation}

For negative $J$, the energy increases with $M_{0}$; for positive
$J$, it is independent of it: it is however never decreasing with
$M_{0}$. This monotonic non-decreasing behavior is in fact a very
general property of any spin-rotation-invariant many-body Hamiltonian,
$\hat{H}$, the ensemble-constrained energy $\mathcal{E}(M_{0})$
is a convex and even function \footnote{Let $\mathcal{E}(\mathbf{M})$ be the lowest ensemble energy attainable
subject to the vector constraint $\int\mathbf{m}\,d^{3}r=\mathbf{M}$.
That $\mathcal{E}(\mathbf{M})$ is convex ($\mathcal{E}(\lambda\mathbf{M}_{1}+\left(1-\lambda\right)\mathbf{M}_{2})\le\lambda\mathcal{E}(\mathbf{M}_{1})+(1-\lambda)\mathcal{E}(\mathbf{M}_{2})$
) follows from the standard property of a ground-state energy constrained
by a linear functional of the state \citep{Valone1980,Lieb1983}.
That $\mathcal{E}(\mathbf{M})$ is even follows from $\hat{H}$'s
invariance to global spin rotation changing $\mathbf{M}$ to $-\mathbf{M}$,
leaving $\mathcal{E}(\mathbf{M})=\mathcal{E}(-\mathbf{M})$. The function
$\mathcal{E}(M_{0})\equiv\mathcal{E}(M_{0}\hat{\mathbf{z}})$ inherits
both properties.}, and hence non-decreasing for $M_{0}\ge0$, with $\mathcal{E}(0)$
the global minimum.

\subsection{Constraining the magnetization in SDFT}\label{subsec:Constraint-magnetiz-SDFT}

To constrain the magnetization to a chosen value $M_{0}\ge0$ we use
the augmented Lagrangian 
\begin{equation}
\mathcal{\mathcal{E}}_{KS}\left(M_{0}\right)=E_{KS}[n,\mathbf{m}]-\lambda\bigl(|\mathbf{M}|-M_{0}\bigr)+\frac{\alpha}{2}\bigl(|\mathbf{M}|-M_{0}\bigr)^{2},\label{eq:aug-lagrangian}
\end{equation}
which combines a Lagrange multiplier $\lambda$ with a quadratic penalty
$\alpha$ that stabilizes self-consistency. Varying with respect to
$\mathbf{m}$ yields Kohn--Sham equations containing a uniform field
$\mathbf{B}_{\lambda}=-\bigl[\lambda-\alpha\bigl(|\mathbf{M}|-M_{0}\bigr)\bigr]\hat{\mathbf{M}}$
along the total-spin direction: 
\begin{align}
\Bigl[-\tfrac{1}{2}\nabla^{2}+v_{s}+\bm{\sigma}\cdot\bigl(\mathbf{B}_{\mathrm{xc}}+\mathbf{B}_{\lambda}\bigr)\Bigr]\varphi_{i} & =\varepsilon_{i}\varphi_{i}.\label{eq:constrained-ks}
\end{align}
The multiplier is updated between cycles by $\lambda\leftarrow\lambda-\alpha(|\mathbf{M}|-M_{0})$.
At convergence $|\mathbf{M}|=M_{0}$, and $\mathcal{E}\left(M_{0}\right)$
equals the KS energy: both constraint and penalty terms vanish. The
field fixing the moment is thus $\mathbf{B}_{\lambda}=-\lambda\hat{\mathbf{M}}$
and the final Lagrange multiplier is the derivative with respect to
the constraint: 
\begin{equation}
\lambda=\mathcal{\mathcal{E}}^{\prime}_{KS}\left(M_{0}\right).\label{eq:Lambda}
\end{equation}
Notes: (1) $\mathcal{\mathcal{E}}_{KS}(M_{0})\ge E_{KS}(M_{\mathrm{gs}})$,
where $M_{\mathrm{gs}}=|\mathbf{M}_{\mathrm{gs}}|$ is the unconstrained
ground-state moment, for which $\lambda=0$; (2) the constraint field
does not exert a net torque \footnote{The uniform magnetic field $\mathbf{B}_{\lambda}$ is parallel to
$\mathbf{M}$, and its net torque $\int\mathbf{m}\times\mathbf{B}_{\lambda}\,d^{3}r=\mathbf{M}\times\mathbf{B}_{\lambda}$
vanishes.}; (3) at the energy minimum the total exchange-correlation torque
vanishes as well.

Since the ensemble \citep{Valone1980,Lieb1983} formulation of $F$
inherits the convexity of $\mathcal{E}\left(\mathbf{M}\right)$ discussed
above, the constrained Kohn--Sham energy $\mathcal{\mathcal{E}}_{KS}\left(M_{0}\right)$
is monotonically non-decreasing, and the Lagrange multiplier $\lambda$
of Eq.~(\ref{eq:Lambda}) is non-negative. An approximate $F$, however,
might not respect convexity and the above may not hold. Indeed we
see violations in both LoC and SoF, for example, the discontinuous
drop at $M_{0}=6$ noted in Sec.~\ref{sec:manganese-dimer}.

\subsection{The spin according to the Kohn--Sham determinant}

The energy fit of Sec.~\ref{sec:manganese-dimer} compares $\mathcal{E}_{KS}\left(M_{0}\right)$
with the exact constrained minimum of the model, Eq.~(\ref{eq:minHeisenbergE}).
Independently of that fit, the Kohn--Sham determinant itself can
be interrogated for the local moment it implies. This uses the determinant's
own $\left\langle \hat{\mathbf{S}}^{2}\right\rangle _{\Phi}\equiv\left\langle \Phi\left|\hat{\mathbf{S}}^{2}\right|\Phi\right\rangle $,
which is not the interacting value $\left\langle \Psi\left|\hat{\mathbf{S}}^{2}\right|\Psi\right\rangle $
entering Eq.~(\ref{eq:affine}). We model the determinant as a product
state $\left|\Phi\right\rangle $ of the two sites, for which $\left\langle \hat{\mathbf{S}}_{1}\cdot\hat{\mathbf{S}}_{2}\right\rangle _{\Phi}=\left\langle \hat{\mathbf{S}}_{1}\right\rangle _{\Phi}\cdot\left\langle \hat{\mathbf{S}}_{2}\right\rangle _{\Phi}$
and thus 
\begin{align}
\left\langle \hat{\mathbf{S}}^{2}\right\rangle _{\Phi} & =2s\left(s+1\right)+2\left\langle \hat{\mathbf{S}}_{1}\right\rangle _{\Phi}\cdot\left\langle \hat{\mathbf{S}}_{2}\right\rangle _{\Phi}.
\end{align}
We take $\left\langle \hat{\mathbf{S}}_{1}\right\rangle _{\Phi}$
and $\left\langle \hat{\mathbf{S}}_{2}\right\rangle _{\Phi}$ to be
two vectors of length $\bar{s}$ with an angle $\alpha$ between them,
so 
\begin{align}
\left\langle \hat{\mathbf{S}}_{1}\right\rangle _{\Phi}\cdot\left\langle \hat{\mathbf{S}}_{2}\right\rangle _{\Phi} & =\bar{s}^{2}\cos\alpha.
\end{align}
The length of the total moment, $\left|\left\langle \hat{\mathbf{S}}_{1}\right\rangle _{\Phi}+\left\langle \hat{\mathbf{S}}_{2}\right\rangle _{\Phi}\right|=2\bar{s}\cos\left(\alpha/2\right)$,
is fixed by the constraint to $M_{0}/2\equiv\sqrt{x}$. Hence $x=2\bar{s}^{2}\left(1+\cos\alpha\right)$,
$2\left\langle \hat{\mathbf{S}}_{1}\right\rangle _{\Phi}\cdot\left\langle \hat{\mathbf{S}}_{2}\right\rangle _{\Phi}=x-2\bar{s}^{2}$,
and 
\begin{equation}
\left\langle \hat{\mathbf{S}}^{2}\right\rangle _{\Phi}=x+c,\label{eq:S2KS}
\end{equation}
a linear function of $x$ of slope one and intercept $c=2\left(s\left(s+1\right)-\bar{s}^{2}\right)$.
Two similar quantities appear here. The first is $s$, the parameter
of the Heisenberg model, the size of each local spin. The second,
$\bar{s}=|\langle\hat{\mathbf{S}}_{a}\rangle|$, is the length the
Kohn--Sham determinant assigns to the local spin moment on each atom.
The determinant is not the interacting wave function, so it need not
give $\bar{s}=s$. However, since the expectation value of a spin-$s$
operator cannot exceed $s$ in magnitude, we expect $\bar{s}\le s$
and therefore $c\ge2s$, with equality only when each site is maximally
polarized. When these inequalities are violated, the local-moment
picture is itself in doubt.

\section{Orientation-relaxed self-consistency}\label{sec:Orientation-relaxed-self-consist}

Because $E_{\mathrm{xc}}$ of Eq.~(\ref{eq:SoF-Exc-definition})
is not invariant under global spin rotations, a converged Kohn--Sham
solution may not have the optimal overall spin orientation. A rotation
\emph{antiparallel} to $\bm{\mathcal{T}}=\bm{\mathcal{T}}_{\mathrm{xc}}+\bm{\mathcal{T}}_{\mathrm{ext}}$
lowers the energy\footnote{An infinitesimal rotation of the magnetization with a rotation vector
$\delta\bm{\omega}$ is $\delta\mathbf{m}=\delta\bm{\omega}\times\mathbf{m}$
(Eq.~(\ref{eq:infinitesimal-rotation})). In the presence of a magnetic
field \textbf{$\mathbf{B}=\mathbf{B}_{\mathrm{ext}}+\mathbf{B}_{\mathrm{xc}}$}
this rotation changes the energy by $\delta E=\int\mathbf{B}\cdot\delta\mathbf{m}\,d^{3}r=\delta\bm{\omega}\cdot\int\mathbf{m}\times\mathbf{B}\,d^{3}r=\delta\bm{\omega}\cdot\bm{\mathcal{T}}$.}. We therefore alternate SCF convergence with rigid spin rotations
until the total torque vanishes. Given a converged solution with torque
$\bm{\mathcal{T}}\neq\mathbf{0}$, we rotate all occupied Kohn--Sham
spinors $\psi_{n}$ by $\bm{\omega}=-\eta\bm{\mathcal{T}}$ with a
small step $\eta>0$, using the one-electron rotation $\hat{U}_{\bm{\omega}}=e^{-i(\bm{\omega}\cdot\bm{\sigma})/2}=\cos\tfrac{\omega}{2}\,\mathbb{I}_{2}-i\sin\tfrac{\omega}{2}\,(\hat{\mathbf{n}}\cdot\bm{\sigma})$,
$\psi^{\prime}_{n}=\hat{U}_{\bm{\omega}}\psi_{n}$. This rotates the
magnetization rigidly, 
\begin{equation}
\mathbf{m}'=\sum_{n}\psi^{\prime\dagger}_{n}\bm{\sigma}\psi^{\prime}_{n}=\sum_{n}\psi^{\dagger}_{n}\bigl(\hat{U}^{\dagger}_{\bm{\omega}}\bm{\sigma}\hat{U}_{\bm{\omega}}\bigr)\psi_{n}=\mathcal{R}_{\bm{\omega}}\mathbf{m},
\end{equation}
where the middle equality is Eq.~(\ref{eq:Pauli-sigma-under-spin-rotation}).
The rotated spinors seed a new SCF cycle, and the two steps are repeated
until $\bm{\mathcal{T}}=\mathbf{0}$ to the convergence threshold.
The same rotation, applied once with a prescribed $\bm{\omega}$ and
without re-converging, generates the rigid-rotation scans of Sec.~\ref{sec:manganese-dimer}.

\section{The current-density route}\label{sec:cdft-route}

 Capelle and Gross \citep{CapelleGross1997} appear to obtain what
$\perp$SDFT cannot: a source-free field for the exchange-correlation
energy of SDFT itself, with no change of functional. Their starting
point is the same partial integration as in Ref.~\citep{Sharma2018},
$\int\mathbf{B}\cdot\mathbf{m}=\int\mathbf{A}\cdot(\nabla\times\mathbf{m})$
for a finite system, but they carry it into current-density functional
theory (CDFT), whose variables include the orbital current. The vector
potential then couples to the total current $\mathbf{j}_{m}=\mathbf{j}_{{\rm p}}+\nabla\times\mathbf{m}$
(in our units), and Levy--Lieb constrained search gives $F$, $T_{{\rm s}}$
and $E_{{\rm xc}}$ as functionals of $(n,\mathbf{j}_{m})$. On the
set $\mathcal{M}=\{(n,\mathbf{m})\mid\mathbf{j}_{{\rm p}}[n,\mathbf{m}]=0,\ \mathbf{B}_{{\rm ext}}[n,\mathbf{m}]=0\}$
of zero-field ground states without orbital current, the SDFT and
CDFT ground states coincide, so $E_{{\rm xc}}[n,\mathbf{m}]=E^{{\rm CDFT}}_{{\rm xc}}[n,\nabla\times\mathbf{m}]$
and $\mathbf{B}_{{\rm xc}}=\nabla\times\mathbf{A}^{{\rm CDFT}}_{{\rm xc}}$
there; and since gauge invariance makes a CDFT functional depend on
the current only through its vorticity \citep{VignaleRasolt1988},
$E_{{\rm xc}}[n,\mathbf{m}]=\bar{E}_{{\rm xc}}\big[n,\nabla\times\big(\tfrac{\nabla\times\mathbf{m}}{n}\big)\big]$
on $\mathcal{M}$. Adding a gradient to $\mathbf{m}$ does not change
the right-hand side, from which Ref.~\citep{CapelleGross1997} concludes
that $\mathbf{B}_{{\rm xc}}$ has no longitudinal part. The claim
holds on $\mathcal{M}$ but cannot be extended. Adding $\nabla\chi$
to $\mathbf{m}$ assumes that $\left(n,\mathbf{m}+\nabla\chi\right)$
still belongs to $\mathcal{M}$. However, it need not. Independently
of membership, $\bar{E}_{{\rm xc}}$ is built from $\nabla\times\mathbf{m}$,
which is not covariant under a global spin rotation, Eq.~(\ref{eq:curl-covariance}),
whereas $E_{{\rm xc}}[n,\mathbf{m}]$ is invariant, Eq.~(\ref{eq:Exc-invariance-To-Spin-Rotation}).
Capelle and Gross note themselves that the surface term dropped in
the partial integration confines Eq.~(26) of Ref.~\citep{CapelleGross1997}
to finite systems, and that in the homogeneous limit, where $\nabla\times\mathbf{m}=0$,
the identity would make $E_{{\rm xc}}$ independent of the magnetization
\citep{CapelleGross1997}. $\perp$SDFT avoids this last defect by
keeping the uniform component in $\mathbf{m}_{\perp}$ (Appendix~\ref{sec:TheTransverseProjection}),
as Ref.~\citep{Sharma2018} does by adding the total moment to the
domain of $\tilde{E}_{{\rm xc}}$ for periodic systems, and for the
homogeneous electron gas $\tilde{E}_{{\rm xc}}=E_{{\rm xc}}$by the
criterion of Sec.~\ref{sec:transverse-theory}. In sum, neither route
yields a source-free field for the exchange-correlation energy of
SDFT. The CDFT identity holds on $\mathcal{M}$ and does not extend
beyond it; $\perp$SDFT holds for every source-free external field,
but does so by replacing $E_{{\rm xc}}$ with $\tilde{E}_{{\rm xc}}$.

\section{$N$-representability and the saturated parent}\label{sec:nrep}

\emph{Violation of $N$-representability.} The construction plugs
the transverse part $\mathbf{m}_{\perp}$ into the ``slot'' of $\mathbf{m}$
in the parent functional. This violates the $N$-representability
condition $|\mathbf{m}_{\perp}(\mathbf{r})|\le n(\mathbf{r})$ \citep{Gontier2013},
because $\mathbf{m}_{\perp}$ decays algebraically, Eq.~(\ref{eq:dipole-field})
of Appendix~\ref{sec:TheTransverseProjection}, while $n$ decays
exponentially for finite systems. The defect belongs to the bare-parent
approximation, not to $\perp$SDFT: $\tilde{E}_{\mathrm{xc}}$ is
defined on the transverse parts of $N$-representable magnetizations
and asks nothing of $|\mathbf{m}_{\perp}|$, whereas the parent was
built for a magnetization that satisfies the bound.

We handle this by extending $E^{\mathrm{parent}}_{\mathrm{xc}}$ beyond
its physical domain, evaluating it at the saturated polarization $\min(|\mathbf{m}_{\perp}|,n)$.
So we extend the parent, not the projection: the saturated parent
is still a functional of $n$ and \textbf{m}, so Eq.~(\ref{eq:Bxc-general})
and $\nabla\cdot\mathbf{B}_{\mathrm{xc}}=0$ are intact. In spatial
domains where $|\mathbf{m}_{\perp}|>n$, $\varepsilon_{\mathrm{xc}}$
stops depending on $\mathbf{m}_{\perp}$, and $\mathbf{B}^{\mathrm{parent}}_{\mathrm{xc}}$
vanishes there: it is active only at points where $|\mathbf{m}_{\perp}|<n$.
This does not mean $\mathbf{B}_{\mathrm{xc}}$ vanishes in the saturated
region: the projection in Eq.~(\ref{eq:Bxc-general}) is nonlocal,
so the field there is inherited from the points where the parent is
active.

\end{document}